\documentclass[journal=jacsat]{achemso}
\usepackage[version=3]{mhchem}
\usepackage{etoc} 
\usepackage{color}
\usepackage{units}
\usepackage{float}
\usepackage{ulem}
\usepackage{amsmath}

\definecolor{comment1}{rgb}{0,0,0}
\newcommand{\RV}[1]{{\color{comment1} #1}}

\author{Aaron R. Finney}
\email{a.finney@ucl.ac.uk}
\author{Matteo Salvalaglio}
\email{m.salvalaglio@ucl.ac.uk}
\affiliation[UCL]
{Thomas Young Centre and Department of Chemical Engineering, University College London, London WC1E 7JE, United Kingdom}

\title{Properties of aqueous electrolyte solutions at carbon electrodes: effects of concentration and surface charge on solution structure, ion clustering and thermodynamics in the electric double layer}

\begin{document}
\etocdepthtag.toc{mtchapter}
\etocsettagdepth{mtchapter}{subsection}
\etocsettagdepth{mtappendix}{none}

\singlespacing
\begin{abstract}
\noindent Surfaces are able to control physical-chemical processes in multi-component solution systems and, as such, find application in a wide range of technological devices. Understanding the structure, dynamics and thermodynamics of non-ideal solutions at surfaces, however, is particularly challenging. Here, we use Constant Chemical Potential Molecular Dynamics (C$\mu$MD) simulations to gain insight into aqueous NaCl solutions in contact with graphite surfaces at high concentrations and under the effect of applied surface charges: conditions where mean-field theories describing interfaces cannot (typically) be reliably applied.
We discover an asymmetric effect of surface charge on the electric double layer structure and resulting thermodynamic properties, which can be explained by considering the affinity of the surface for cations and anions and the cooperative adsorption of ions that occurs at higher concentrations.
We characterise how the sign of the surface charge affects ion densities and water structure in the double layer and how the capacitance of the interface---a function of the electric potential drop across the double layer---is largely insensitive to the bulk solution concentration.
Notably, we find that negatively charged graphite surfaces induce an increase in the size and concentration of extended liquid-like ion clusters confined to the double layer.
Finally, we discuss how concentration and surface charge affect the activity coefficients of ions and water at the interface, demonstrating how electric fields in this region should be explicitly considered when characterising the thermodynamics of both solute and solvent at the solid/liquid interface.

\end{abstract}

\section{Introduction}
\label{sec:introduction}
Carbon-electrolyte interfaces often feature in technologies and devices designed for energy storage \cite{frackowiak_carbon_2001,wang_electrochemical_2016,simon_materials_2008} and water desalination\cite{porada_review_2013,cohen-tanugi_water_2012}. 
Moreover, carbon allotropes are increasingly employed as nano-reactors,\cite{tian_nanoengineering_2019} as well as supports for liquid-phase catalysts\cite{julkapli_graphene_2015}. 
A molecular-level picture of the structure and dynamics of multi-component liquid phases at the carbon interface is important to understand the physical chemistry involved in such technologies/devices in order to improve their design for functional applications.
Molecular simulations, particularly molecular dynamics (MD), provide powerful tools to investigate such systems at the atomic level.\cite{elliott_electrochemical_2022}
By explicitly capturing the atomistic details of the solid/liquid interface, MD-based methods enable predictions regarding the effect of changes to the bulk solution composition and the applied interfacial potential (that gives rise to a surface charge) on the properties of the so-called electric double layer (EDL). 
In turn, this allows for an assessment of the suitability of mean-field models that are commonly used to describe and predict the structure and electrochemical properties of solid-solution interfaces.\cite{petsev_molecular_2021}

Gouy-Chapman theory predicts a monotonically decreasing concentration of ions in the immediate vicinity of electrodes with the same sign of charge, while the concentration of ions with opposite charge to the surface increases smoothly according to a Boltzmann distribution; thus, the solution screens the surface charge by establishing a diffuse EDL. \cite{bard_electrochemical_2001}
This fundamental model for the structure of charge carriers at electrodes inadequately describes the EDL when large potentials are applied and in the presence of high electrolyte concentrations. 
By neglecting ion finite-sizes and their correlations, it fails to explain the change in the electrical properties of the graphite-electrolyte interface due to specific ion effects, which was demonstrated across the series of alkali chlorides at graphite. \cite{iamprasertkun_capacitance_2019,zhan_specific_2019}
The simple picture of the EDL was was later developed to address some of these shortcomings, by accounting for the specific adsorption of ions at the electrode and the role that ion solvation spheres play in defining the inner- and outer-Helmholtz plane.\cite{grahame_electrical_1947} 
In this framework, the solution-side of the EDL is modelled as a series of plate capacitors; nonetheless, it is assumed that the finite size of charge carriers can be ignored in the diffuse layer.
At low concentrations of simple salts---such as NaCl---in water, these simple mean-field-based models were suggested to provide a reasonable approximation of the EDL structure, \cite{fedorov_ionic_2014} especially as charge transfer between the electrode and charge carriers in solution is low. \cite{zhan_specific_2019}
However, the combination of high salt concentrations and large surface charge densities results in conditions where the solvation, finite size and cooperative adsorption of ions cannot be neglected. More sophisticated mean-field models of the EDL were developed to address some of these effects. \cite{borukhov_steric_1997,goodwin_mean-field_2017,yin_mean-field_2018,uematsu_effects_2018,hedley_dramatic_2023,mceldrew_ion_2021,goodwin_gelation_2022}

Our recent simulations demonstrated how asymmetric electrolyte adsorption gives rise to alternating cation and anion-rich aqueous solution layers perpendicular to planar graphene and graphite substrates at moderate-to-high alkali chloride solution concentrations ($\sim1$ M and above). \cite{finney_electrochemistry_2021,pasquale_constant_2022} 
This behaviour is due to the partial saturation of ions in solution layers in contact with the surface that emerges in the EDL. \cite{elliott_qmmd_2020,dockal_molecular_2019,dockal_molecular_2022,pasquale_constant_2022,elliott_electrochemical_2022}
This picture of the EDL is reminiscent of the structures observed in ionic liquids at charged surfaces and requires a treatment of the EDL that accounts for the finite size of charge carriers accumulating at the interface.\cite{kornyshev_double-layer_2007,fedorov_ionic_2008,fedorov_ionic_2014}
The asymmetric ordering of ions results in charge fluctuations in this region---typically four-to-five liquid layers deep---and a departure from descriptions of the EDL expected from the established mean-field models described above.\cite{schmickler_interfacial_1996,petsev_molecular_2021}

Thanks to the adoption of the Constant Chemical Potential Molecular Dynamics (C$\mu$MD) method,\cite{perego_molecular_2015,karmakar_non-equilibrium_2023} which maintains a constant thermodynamic driving force associated with ion adsorption, we were able to quantify the electric potential drop across the EDL and the excess chemical potential for ions at the solid-solution interface,\cite{finney_electrochemistry_2021,finney_bridging_2022} 
In C$\mu$MD, the use of an explicit molecular reservoir coupled to the model interface prevents any ion depletion in the bulk solution, which would otherwise occur in typical finite-sized MD simulations when ions adsorb at an interface. 

Here, we extend our analysis to consider concentrated NaCl(aq) solutions in contact with charged graphite and the resulting properties of the solution side of the EDL.
In our analysis of the simulation results, we pay particular attention to the thermodynamic and structural properties of the solvent (as well as ion speciation).
Understanding how the presence of ions and surface charge control the thermodynamics of solvent is essential to predict the activity of interfaces for applications in catalysis, and recent computational studies have demonstrated how interfaces impact the ability of the aqueous medium to screen Coulombic interactions due to a changing dielectric constant. \cite{olivieri_confined_2021}

In what follows, we recap the effect of concentration on the structure and properties of ions in solution at neutral graphite before considering the combined effects of concentration and surface charge.
We characterise the structural properties of water molecules in the EDL when compared to bulk solutions, pure liquid water and ice.
Finally, we evaluate the electrical properties of the EDL and use this information to calculate how the activity constants for ions and water change on moving from the bulk solution towards the graphite surface.

\section{Computational Methods}
\label{sec:methods}
Following the protocol proposed by Finney et al. \cite{finney_electrochemistry_2021}, all simulations were performed using the Joung and Cheatham\cite{joung_determination_2008} force field to describe the interactions of ions with SPC/E water\cite{berendsen_missing_1987}. 
Graphite was modelled using the OPLS/AA force field,\cite{jorgensen_development_1996} while the intermolecular interactions between carbon and water were modelled using pairwise potentials fitted to water adsorption energies obtained via random phase approximation calculations.\cite{ma_adsorption_2011,wu_graphitic_2013}
\RV{Several force fields are available to model the interactions of carbon with water, and comparisons of some of the different models are available in the literature. \cite{werder_water-carbon_2003,wu_graphitic_2013,kim_wetting_2014,li_water_2017}

The carbon-water model adopted here predicts a water contact angle of $\sim 40^\circ$, with small changes to this mean value being dependent upon the number of carbon layers in the substrate and the truncation distance used in the interaction potential. \cite{wu_graphitic_2013}
The contact angle is more acute than that predicted by earlier force fields; however, it was shown in experiments that graphene becomes less hydrophilic when exposed to air and the surface becomes populated by contaminants (hence, a smaller contact angle should be reproduced by the model than was initially thought). \cite{schrader_ultrahigh_1975,schrader_ultrahigh-vacuum_1980,prydatko_contact_2018}
The contact angles for pristine graphene and graphite were found to be $42 \pm 7^\circ$ and $42 \pm 3^\circ$, respectively.\cite{schrader_ultrahigh-vacuum_1980,prydatko_contact_2018}
A recent exhaustive computational study of the interaction energies of water with graphene using quantum mechanical calculations suggest an upper bound to the contact angle of water on graphene of $56^\circ$, as informed by dynamical simulations of coarse-grained water molecules at the carbon surface, where interaction potentials were fitted to the results from calculations at a higher level of theory.\cite{brandenburg_physisorption_2019}
Our model, therefore, captures reasonably well the thermodynamics of water at the carbon interface (as determined by the surface tension); furthermore, it predicts the correct radial breathing mode frequency for carbon nanotubes in water. \cite{wu_graphitic_2013}
}

Ion-carbon interactions were modelled using potentials fitted to the results from electronic structure calculations that capture the polarisability of the carbon surface in the presence of ions surrounded by a conductor-like polarisable continuum, mimicking the presence of a solvent.\cite{williams_effective_2017}
Despite components of the force field being constructed from various sources, it is important to recognise that consistent descriptions of ions and water molecules (i.e., Joung and Cheatham ions and SPC/E water) were used for the fitting of pairwise potentials throughout.

The GROMACS 2018.6\cite{hess_gromacs_2008} MD engine was adopted to perform simulations within the NVT ensemble unless otherwise stated.
Atom positions were evolved during the simulations using a leapfrog time integrator with a 2~fs timestep; as such, water intramolecular degrees of freedom were constrained using the LINCS algorithm. \cite{hess_lincs_1997}
Intermolecular interactions were computed for atoms within 0.9~nm, and long-range electrostatics were treated using smooth particle mesh Ewald summation. \cite{essmann_smooth_1995}
The temperature was held constant at 298~K (within fluctuations) using the Bussi-Donadio-Parrinello thermostat.\cite{bussi_canonical_2007}

The simulation set-up for graphite in contact with NaCl(aq) solutions follows our previous work. \cite{finney_electrochemistry_2021}
This involved preparing an eight-carbon layer $2.7 \times 5.4 \times 5.5$~nm ($x \times y \times z$) graphite slab with basal surfaces perpendicular to the simulation cell $x$-axis. 
1,672 Na$^+$ and Cl$^-$ ions, as well as 13,819 water molecules, were placed in the orthorhombic simulation cell with periodic boundaries in all three dimensions.
With carbon atoms fixed at their lattice positions, a 0.2~ns MD simulation was performed to relax the simulation cell volume in the NPT ensemble using the barostat of Berendsen et al.\cite{berendsen_molecular_1984} at a pressure of 1~atm.
Following this equilibration step, the ions in solution were accumulated in a reservoir region far away from the graphite surface(s) by applying an external harmonic potential to the distance between the surface and ions (using the PLUMED v2.5 plugin\cite{tribello_plumed_2014} with force constant, $k=3 \times 10^5$ kJ mol$^{-1}$).
The minimum distance between carbon atoms and ions in this external bias was 6~nm.
Simulations in the NVT ensemble were performed until the ions were at least 5.9~nm from the carbon slab.
The final configuration of the system from this preparatory phase was taken as the starting structure for 100~ns C$\mu$MD simulations, where the final 50~ns steady-state trajectory window was used in all analyses of the interfacial properties. A similar procedure was used to prepare simulations of graphite in contact with pure water; here, however, no ions were included in the simulation cell, and it was not necessary to prepare the ionic reservoir.

For all C$\mu$MD simulations, PLUMED 2\cite{tribello_plumed_2014} was utilised to compute the external forces required to control the solute density ($n$) in a 2.2~nm control region, whose innermost edge (closest to a graphite basal plane) was $x_F=3.7$~nm from the centre of the simulation cell $x$-axis, defining the origin.
The C$\mu$MD force on ion $i$ takes the functional form,
\begin{equation}
    F_i(x)=\frac{k_i}{4 \omega}(n_i^\mathrm{CR} - n_i^t) \left[ 1 + \cosh \left( {\frac{x-x_F}{\omega}} \right) \right]^{-1}
\end{equation}
where $\omega$ tunes the width of the force region and was taken to be 0.01\% of the cell length in $x$; superscript CR and $t$ indicate the instantaneous and target value of $n$ in the control region; and $k_i=2\times 10^5$~kJ mol$^{-1}$ is the force constant for the function that acts like a semi-permeable membrane for the ions.

Standard MD simulations were performed to simulate graphite in contact with water, which we subsequently refer to as 0 mol dm$^{-3}$ (M).
In the presence of ions, on the other hand, C$\mu$MD simulations were performed where the target ion density was 0.6022, 3.0110 and 6.022~nm$^{-3}$, equating to molar concentrations of 1, 5 and 10 M. 
When the simulations reached a steady state, the concentrations of ions were maintained at $1.2 \pm 0.03$, $5.01 \pm 0.05$ and $9.23 \pm 0.07$ M.
The difference between the target and evaluated concentrations is small and is due to the relative occupancy of the ionic reservoir and the parameters used to apply C$\mu$MD forces; this is not particularly important for the current study, and the simulations can be prepared to ensure that concentrations precisely match the target value, if necessary.
Both 5 and 10 M cases are beyond the solubility for NaCl(s), determined to be approximately 3.5~M for the Joung and Cheatham force field; the solubility of halite is 3.7 mol~kg$^{-1}$,\cite{benavides_consensus_2016} which is approximately 3.5 M according to a $2^\mathrm{nd}$ order polynomial fitting of molarities ($c$) as a function of molalities ($b$) obtained from steady-state bulk NaCl(aq) solutions ranging from 1 to 16 mol~kg$^{-1}$ ($b=-0.0174c^2+0.9822c+0.0537$, where $b$ and $c$ here refer to numeric values ignoring units.).

We explored the effects of applied surface charge in systems at all three sampled bulk solution concentrations as well as in simulations of graphite in contact with pure water.
To achieve this, we applied uniform charges to the outermost carbon atoms in the graphite slab; equal charges with the opposite sign were applied to 1144 carbon atoms on each face of the graphite slab to generate charge densities, $|\sigma| = 0.19$, 0.39, 0.58 and 0.77~$e$~nm$^{-2}$.
(We use the descriptors positively/negatively charged surface and positive/negative electrode interchangeably throughout.)
As such, a single simulation provides information on the effect of positive and negative applied potentials by examining the interface on different sides of the graphite slab.
The total charge in the simulation cell was, therefore, zero.
We believe that this approach to applying charges to the surface is reasonable for the current study; indeed, when we tested a Drude oscillating charge model for the surface charge polarisability at 1 M, we did not observe significant differences in the properties of the interface when compared with the uniform distribution of charge to carbon centres discussed below.
More sophisticated models of the surface charge polarisability have been developed; \cite{elliott_qmmd_2020,coretti_metalwalls_2022} although a comparison of the accuracy and applicability of these methods is beyond the scope of the current work.

\section{Results and Discussion}

\label{sec:results}

\subsection{Solution structure at charged graphite}
\begin{figure}[H]
    \centering
    \includegraphics[width=1\linewidth]{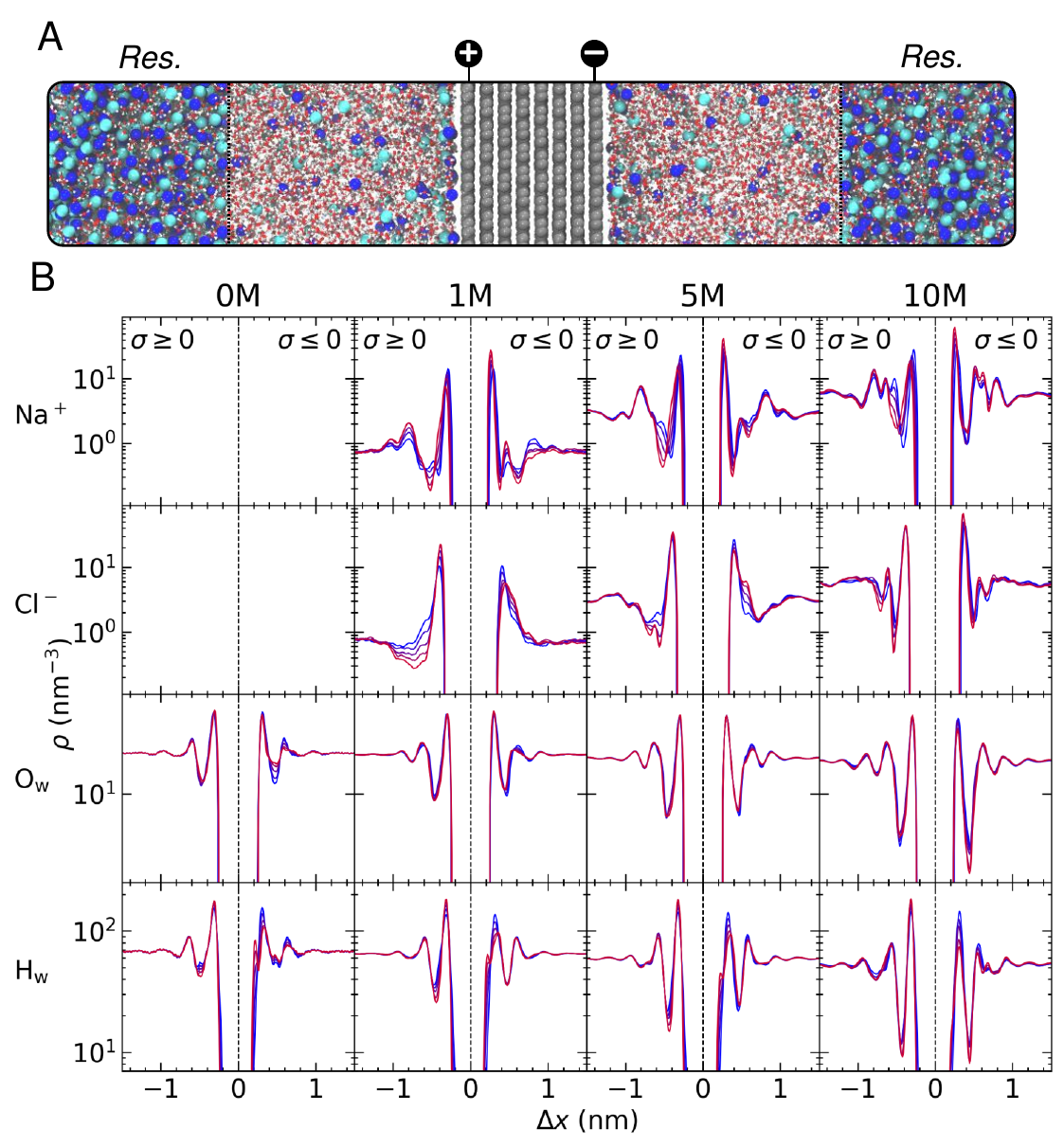}
    \caption{A) Snapshot from a C$\mu$MD simulation of NaCl(aq) at charged graphite projected onto cell $x$ and $y$ axes. The dashed lines indicate the edge of the reservoir (\textit{Res.}); grey, blue, cyan, red and white spheres indicate C, Na$^+$, Cl$^-$, O and H atoms, respectively. B) Mean solution atom densities, $\rho$, determined as a function of distance from the graphite electrode, $\Delta x$, for a range of target bulk solution ion concentrations. The solution concentration and atom type are indicated at the top and left of the grid. Colours blue$\rightarrow$red indicate increasing solution charge densities from $|\sigma | =0 \rightarrow 0.77~$e~nm$^{-2}$. }
    \label{fig:densities}
\end{figure}

In this section, we discuss the salient features of the steady-state structure of NaCl(aq) solutions at graphite surfaces when negative and positive surface charges are applied to carbon atoms, equating to surface charge densities ($\sigma$) in the range $0 - \pm 0.77~$e~nm$^{-2}$.
The target bulk ion solution concentrations in our C$\mu$MD simulations here were 1, 5 and 10 M.
In addition, we also perform simulations of neutral and charged graphite in contact with pure water.
A snapshot of a typical C$\mu$MD simulation is provided in Figure \ref{fig:densities}~A, where the ion-rich reservoir can be seen spanning the periodic boundaries, far from the carbon-electrolyte interface. 
As ions accumulate in the EDL, the solution in contact with the graphite surface is replenished with ions from the reservoir to maintain a constant thermodynamic driving force for the process, ensuring that the solution, several nanometres from the interface---which we refer to as the \textit{bulk}---is electroneutral.
In the following subsections, we focus our analysis on the steady state that emerges between species in solution in this bulk region and the EDL.

\paragraph{Density profiles---concentration and charge effects}
The one-dimensional atom densities of solution species perpendicular to the graphite basal surface are reported in Figure \ref{fig:densities}~B. In addition, Figures \ref{fig:d0M}--\ref{fig:d10M} provide the same densities on a linear scale over a wider range of $\Delta x$.
In line with our previous simulation studies\cite{finney_electrochemistry_2021,finney_bridging_2022} and those from others using different force fields and graphene,\cite{dockal_molecular_2019,dockal_molecular_2022,elliott_qmmd_2020,pasquale_constant_2022} the densities indicate a preference for cation adsorption in the first solution layer above the substrate; this is due to the favourable interactions between positively charged ions and the electron-rich carbon surface (implicitly captured by the force field) \cite{williams_effective_2017}. At the lowest concentrations, we observe a diffuse anion-rich solution layer adjacent to the first cation-rich layer.

As discussed in detail by Finney et al. \cite{finney_electrochemistry_2021}, the asymmetric adsorption gives rise to a surface potential, the magnitude of which is governed by the sharp cation density in $x$ in the first solution layer, even in the absence of an applied surface charge. This effective surface charge is screened by a diffuse anion layer and, at concentrations below 0.6 M, the concentration profile leading to such charge screening is qualitatively consistent with simple mean-field models of the EDL\cite{finney_electrochemistry_2021}
However, at $\sim 1$ M and above, additional cation and anion density peaks are observed in the EDL, and a complex multi-layered solution structure emerges due to the finite size and cooperative adsorption of ions that is explicitly captured by atomistic simulations and is apparent from the blue curves in Figure \ref{fig:densities}~B.
This picture is reminiscent of the structure of ionic liquids at planar surfaces, where the finite size of charge carriers cannot be ignored. \cite{kornyshev_double-layer_2007,fedorov_ionic_2008,fedorov_ionic_2014}
These results highlight the inadequacy of simple mean-field models to predict the structure and, ultimately, the electrochemical properties of the interface for even simple electrolyte solutions at moderate to high ion concentrations. For a more detailed discussion of the solution structure at uncharged graphite, see the discussion by Finney et al. \cite{finney_electrochemistry_2021}

In this work, we focus on the effect of varying graphite surface charges on water and ion atom densities in the EDL. Such variations are reported more clearly in Figure \ref{fig:densdiff}.
In pure water, increasing the surface charge density to $+0.77e$ nm$^{-2}$ results in positive and negative changes to the water oxygen and hydrogen densities ($\rho_{O_w}$ and $\rho_{H_w}$), respectively, in the first solution layer adjacent to the graphite basal surface.
Essentially, the surface charge induces an increased ordering of water molecules locally in the vicinity of the positively charged carbon surface.
At the negative electrode, however, we observe a restructuring of the liquid in the first two water layers as the magnitude of the surface charge, $\sigma$, increases. 
This can be explained by water molecules reorienting to increase the interactions between H-atoms, bearing a positive partial atomic charge, and the excess negative charge uniformly distributed amongst carbon atoms in the outermost graphite layer. The reorientation of water molecules manifests in the density profiles as a splitting of the first $\rho_{H_w}$ peak into two peaks separated by $\sim 0.1$ nm.
The electrostatic repulsion of water oxygen atoms with the surface also displaces molecules in the first liquid layer, as shown by a decrease in $\rho_{O_w}$ in the first peak and a shallower minimum in density between the first two water layers.
Regardless of the sign or magnitude of $\sigma$, perturbations to the liquid structure encompass approximately three water layers, up to $\sim 1$~nm from the graphite basal plane, consistent with other studies of carbon-solution interfaces. \cite{elliott_qmmd_2020,olivieri_confined_2021}

Interestingly, as ions are added to the system, the amplitude for the fluctuations in water density at the interface somewhat diminish, although perturbations to the water structure cf. the bulk extend further into the solution as the magnitude of the applied surface potential increases.
This can be explained by the screening of the surface charge due to the ions accumulating in the EDL.
For example, at 10 M, the ordering of water is observed four-to-five water layers from the surface, and the spacing between peak centres in $\rho_{H_w}$ decrease, most notably when $\sigma$ is negative, due to the complex ion layering that is found under these conditions at the interface.
This can be reconciled by considering how the surface displaces ions with associated water molecules in their solvation spheres---cations, in particular, have a relatively strong solvent coordination sphere, and so any change to the steady-state structure of $\rho_{Na}$ will affect the density of water molecules in the EDL.

Figures \ref{fig:densities}~B and \ref{fig:d10M} show that a single peak in $\rho_{H_w}$ is observed in the first solution layer at all values of $\sigma$ at 10 M, denoting an inhibition of the water structuring found in the absence of electrolyte. 
Moreover, a large cation density in the first solution layer gives rise to a large increase in the local anion density, which is not apparent at 1 M; hence, the cooperative accumulation of ions displaces water in the second solution layer.
Furthermore, the screening of the surface charge by ions at high concentrations mitigates any reorientation of water dipoles.
Figure \ref{fig:densdiff} indicates that at low concentrations, the density of water in the first solution layer increases at the negatively charged surface due to an increased cation density; however, at 10 M, the density change is negative, demonstrating the complex restructuring of water that occurs in the EDL at high solution concentrations.

The perturbation of the solution structure under the effect of increasing surface charge is analogous to an `accordion-like' deformation, where solution layers are compressed under the action of the additional Coulombic forces.
This compression of the solution layers gives rise to further perturbations to the solution structure, as increased ion ordering propagates the effect of the surface charge into solution: a radically different picture of the EDL than those predicted by simple Poisson-Boltzmann-based models.
This feature is also observed as a function of concentration, where partial saturation of the solution with ions in the first layers above the surface result in further deviations from the bulk structure moving away from the interface (compare the blue curves for ion density in Figure \ref{fig:densities}~B).

In general, we conclude that increasing the solution concentration and surface charge have analogous effects on the solution structure in the EDL.
While it is possible to capture the effects of asymmetric adsorption and ion correlations in mean-field models of the EDL\cite{borukhov_steric_1997,goodwin_mean-field_2017,yin_mean-field_2018,uematsu_effects_2018} and the role that water structuring can play in screening the surface potential,\cite{hedley_dramatic_2023}
the complex solution structure observed at the high concentrations and surface charges here suggests that an explicit model for atoms in the EDL is necessary to capture these cooperative, emergent effects.

\paragraph{EDL relaxation in the presence of surface charge}
In order to study the collective dynamics of ionic species at the interface, it is useful to consider how the application of a surface charge changes the composition in the EDL as a function of time.
As such, we performed five additional simulations at 1, 5 and 10 M, where the initial configurations were taken from 50, 60, 70, 80 and 90~ns time points in simulations where the graphite had no applied charge.
In these simulations, however, we set $\sigma$ to $\pm 0.77 \,e$~nm$^{-2}$ on opposite surfaces of the graphite slab.
By starting from an equilibrated steady-state structure in the absence of surface charge, these simulations allow us to investigate how the EDL evolves in time when a surface charge is instantaneously applied.

\begin{figure}[H]
    \centering
    \includegraphics[width=0.85\linewidth]{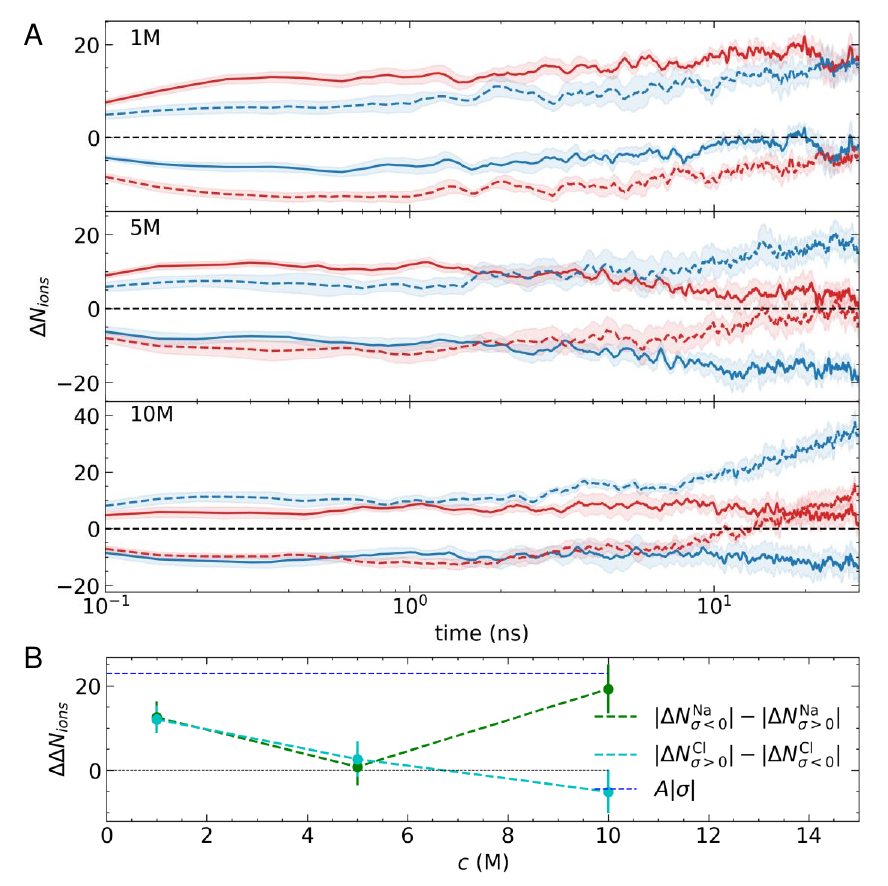}
    \caption{A) Changes to the mean number of ions ($\Delta N_{ions}$) in the EDL (defined as a 2.5 nm region adjacent to the carbon basal surface) when a surface charge is switched on with $\sigma = \pm 0.77 \,e$~nm$^{-2}$. Solid and dashed lines pertain to positively and negatively charged surfaces, and the blue and red colours indicate cations and anions, respectively. The data are averages from five independent simulations with the uncertainties calculated as the standard error of the mean in the data, as shown by the shaded regions. B) A summary of the data in panel A, where the green and cyan points provide the differences between $\Delta N^{Na}$ and $\Delta N^{Cl}$ at the counter and co-electrode in the EDL. Dashed lines provide a linear interpolation between the data as a guide, while the blue dashed line provides the magnitude of the surface charge in units of $e$.
    }
    \label{fig:kinetics}
\end{figure}

Figure \ref{fig:kinetics}~A shows the average change in the number of cations and anions within 2.5~nm from the outermost carbon layer of the graphite surface.
At all simulated concentrations, there is a rapid change in the number of ions in the EDL ($\Delta N_{ions}$).
Indeed, on the log scale provided, the change in $\Delta N_{ions}$ from zero is not apparent, as this occurs during the first $0.05$~ns.
The change is, therefore, extremely rapid as ions are displaced to minimise electrostatic forces.
This behaviour might be expected for this system which is often adopted as a model system to study electric double-layer capacitors (EDLCs).\cite{frackowiak_carbon_2001,wang_electrochemical_2016}
EDLCs offer a high power density, able to deliver and absorb electrical energy at a much higher rate than typical batteries through rapid charge/discharge cycles. 
The excellent cycling capability of EDLCS---which typically undergo millions of charge-discharge cycles (instigated by changing the applied potential)---is possible without significant degradation of the interface, making them attractive options for energy storage devices. \cite{funabashi_introduction_2016}

Despite the rapid change in $\Delta N_{ions}$, it is clear from Figure \ref{fig:kinetics}~A that a longer-timescale relaxation of the EDL composition develops over tens of nanoseconds.
At 1 M, the preference for cations to adsorb in the first solution layer means that there is an asymmetry in the displacement of Cl$^-$ at the negative electrode when compared with Na$^+$ at the positive electrode.
This means that after 30~ns, $\Delta N^{Na}$ increases on both electrodes and so does $\Delta N^{Cl}$, such that $\Delta N^{Na}_{\sigma>0} \approx 0$.
At all stages throughout the relaxation, the surface charge is screened by ions in the EDL ($\Delta N^{Na}_{\sigma<0}-\Delta N^{Cl}_{\sigma<0} \approx \Delta N^{Cl}_{\sigma>0}-\Delta N^{Na}_{\sigma>0} \approx |\sigma|$).

A different behaviour is observed as the concentration of ions is increased. 
At 5 M, Figure \ref{fig:kinetics}~A shows that after 30~ns, the surface charge is almost completely screened by incorporation of cations or their removal from the EDL, whereas $\Delta N^{Cl} \approx =0$, regardless of the sign of the applied potential.
When the concentration is increased to 10 M, $\Delta N^{Na}_{\sigma<0}$ increases and $\Delta N^{Cl}_{\sigma<0}$ is positive; essentially, as more cations are accumulated in the EDL beyond the number necessary to screen the surface charge, ion-ion correlations also induce an increase to the number of anions in the EDL.
Irrespective of this change in behaviour, the surface charge remains screened by changes to ion concentrations in the EDL throughout the period of relaxation.

Figure \ref{fig:kinetics}~B summarises the changes to the EDL ion concentrations. 
Here, we present the differences in absolute changes to the ion concentrations at the counter- (where the sign of the ion charge is opposite to the surface charge) and co-electrodes (where the sign of the charge on ions matches that of the surface charge) for Na$^+$ and Cl$^-$ as $|\Delta N^{Na}_{\sigma<0}|-|\Delta N^{Na}_{\sigma>0}|$ and $|\Delta N^{Cl}_{\sigma>0}|-|\Delta N^{Cl}_{\sigma<0}|$.
The data indicate the affinity of the charged surfaces for cations or anions.
A value of zero in Figure \ref{fig:kinetics}~B indicates equivalent displacement of the ions at the positive and negative electrode, which is the case only when $c$(NaCl) is 5 M.
At 1 M, on the other hand, the surface can accumulate a net excess of ions in equal amounts at both positive and negative $\sigma$.
Finally, at 10 M, the net accumulation of cations exceeds that of anions when comparing surfaces with opposite charges.
Together, these data highlight the multifaceted relaxation of the EDL structure due to asymmetric ion effects and cooperative changes that result, which are typically neglected in analytical models of the EDL.

\subsection{Ion association in solution}
In our previous work, we demonstrated how correlations in bulk solutions give rise to liquid-like NaCl assemblies, also referred to as clusters (across the concentration range sampled in this work), that can reach substantial sizes at high concentrations.\cite{finney_multiple_2022}
\RV{Essentially, NaCl(aq) solutions become increasingly non-ideal as the solution concentration is increased.
This results in small ion associates containing up to three or four ions at $\sim~1$~M, but these clusters can encompass hundreds of ions at the high end of solution concentration, i.e., $\sim~10$~M.

Experimental studies of levitated droplets recently found large liquid-like NaCl clusters at high supersaturations in NaCl(aq), with MD simulation results---using the same force field as the one adopted here---supporting the experimental observations.\cite{hwang_hydration_2021}
The authors speculated that these clusters could play a role in NaCl crystallisation, which was later shown to be the case in simulations of high concentration metastable solutions from our own studies of homogeneous NaCl(aq),\cite{finney_multiple_2022} and in solutions at and beyond the limit of solution stability.\cite{lanaro_birth_2016,jiang_nucleation_2019,bulutoglu_investigation_2022}
Cutting-edge experiments have also demonstrated that NaCl crystals can emerge from disordered ion associates confined to aminated conical carbon nanotubes.\cite{nakamuro_capturing_2021}
In addition, amorphous NaCl solids have been isolated using supersonic spray-drying techniques, where the rapid removal of water from dense ion assemblies occurs before crystal nucleation can occur.\cite{amstad_production_2015}

It was recently shown, using a sophisticated machine learning force field trained on \textit{ab-initio} MD simulation trajectories, that the final stage during NaCl dissolution involves the dissipation of amorphous ion clusters, \cite{oneill_crumbling_2022} and MD simulations using the Joung-Cheatham force field also indicate this mechanism for cluster dissolution. \cite{lanaro_molecular_2015,lanaro_birth_2016}
These studies combined, therefore, suggest that disordered ion assemblies are potentially involved in both the formation and dissolution of solid NaCl and that the Joung-Cheatham force field can capture these mechanisms reasonably well.}
In the presence of graphite, we demonstrated how disordered ion assemblies are stabilised in the EDL due to the increased ion densities in this region and postulated that by catalysing the formation of these clusters, surfaces might control the pathway for NaCl crystallisation. \cite{finney_electrochemistry_2021}
In this section, we explore how applied surface charges affect liquid-like ion assemblies in the EDL.

\begin{figure}[H]
    \centering
    \includegraphics[width=0.9\linewidth]{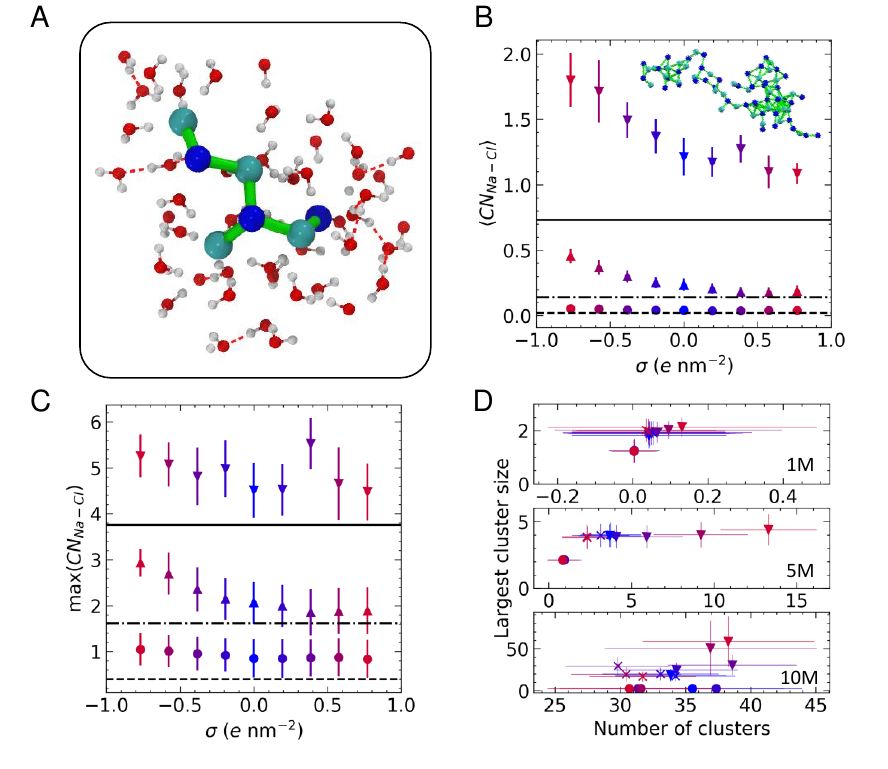}
    \caption{A) A [Na$_3$Cl$_4$]$^{-1}$ cluster at the carbon surface surrounded by its solvation sphere. Na$^+$, Cl$^-$, O and H atoms are shown by blue, cyan, red and white spheres, respectively. Green lines highlight ion connections within 0.4 nm, and H-bonds are indicated by the red dashed lines. B) Average Na-Cl coordination number, $\langle CN_{Na-Cl} \rangle$, in the EDL as a function of surface charge density, $\sigma$. B) Average maximum (max) $CN_{Na-Cl}$ in the EDL. In B and C, $\bigcirc$, $\bigtriangleup$ and $\bigtriangledown$ symbols indicate data for 1, 5 and 10 M with dashed, dotted-dashed and solid lines highlighting the bulk values at 1, 5 and 10 M. D) Number of clusters and largest cluster size in the EDL at the positive ($\times$) and negative electrode ($\bigtriangledown$) with the bulk ($\bigcirc$) values also provided.}
    \label{fig:clusters}
\end{figure}

We determined the connectivity between ions in their first coordination sphere using a truncation distance based on the first minimum in the radial distribution functions between pairs of atoms ($\sim 0.35$~nm) and a continuous rational switching function to smoothly decorrelate ions according to this definition (details are provided in the files that can be obtained by the link in the Data Availability statement
below).
Figure \ref{fig:clusters}~A shows a typical cluster residing in the first solution layer at an uncharged graphite surface when $c$(NaCl) is 5 M.
These clusters evolve their topology over $\sim$ps timescales due to density fluctuations in solution. \cite{finney_multiple_2022}

Figure \ref{fig:clusters}~B provides the average cation-anion coordination number, $\left< CN_{Na-Cl} \right>$, in the EDL when compared to the bulk values.
At 1 M, there is no clear surface effect; however, as the bulk solution concentration increases, the coordination of ions in the EDL exceeds the bulk, in line with our previous observations. \cite{finney_electrochemistry_2021}
When a surface charge is applied, Figure \ref{fig:clusters}~B shows a clear bias for higher levels of ion coordination on the negative electrode. 
At 5 and 10 M, there is a monotonic increase in the average coordination number when $\sigma<0$, while this remains roughly constant when $\sigma>0$, matching bulk values at 5 M, but increasing compared to the bulk at 10 M.

The asymmetric ion coordination can be reasoned by considering the changes in the EDL atom densities as a function of $\sigma$, as provided in Figure \ref{fig:densdiff}.
The affinity for cations to adsorb in the first solution layer increases as negative charges are applied to carbon atoms, and this results in increasing $\rho_{Cl}$ in the vicinity of the surface, particularly as the bulk solution concentration is increased.
This effect is most apparent at 10 M, where the first peaks in \textit{both} $\rho_{Na}$ and $\rho_{Cl}$ increase substantially, in contrast to the positively charged surface, where the increase in the first $\rho_{Cl}$ peak is a small fraction of that on the negative electrode and $\Delta \rho_{Na}$ for the first peak is negative.
Because the highest atom densities occur close to the surface in the EDL, increasing the density in these regions means that the distribution of ions at the negative electrode is more disproportionate than at the positive electrode (see also Figure \ref{fig:d10M}), facilitating greater ion coordination close to the surface.

Figure \ref{fig:clusters}~C provides the maximum cation-anion coordination number: max($CN_{Na-Cl}$).
The plot indicates that there is a greater propensity to form ion pairs in the EDL than in the bulk solution, but there is no clear surface charge effect at the lowest concentration.
Two-coordinate cations are most likely to be observed at 5 M, and this increases to three-fold cation-anion coordination in the EDL when $\sigma \ll 0$, indicating a change from linear ion coordination to branched coordination, consistent with changes to the structure of clusters that are found on increasing ion concentrations at uncharged graphite. \cite{finney_electrochemistry_2021}
The complex, multi-layered solution structure emerging at 10 M means that very high levels of ion coordination are observed in the EDL, irrespective of the sign of the applied charge.

A value of $CN_{Na-Cl}=6$ is consistent with the levels of coordination in the rock salt crystal structure.
The fact that the max($CN_{Na-Cl}$) values approach this limit in the EDL at 10 M supports the hypothesis that crystallisation is promoted in this region, with order emerging from the liquid-like clusters.
Indeed, when $\sigma=+0.58 \,e$ nm$^{-2}$, a high ion density, anhydrous region of the extended cluster could potentially progress to a close-packed crystal structure.
The solutions are highly metastable at this bulk concentration (using the adopted force field, the bulk solutions are metastable at $3.7-15$ mol kg$^{-1}$ \cite{benavides_consensus_2016,jiang_nucleation_2019} which is approximately $3.5-10.9$ M).
Nonetheless, crystal nucleation is a rare event that is unlikely to occur over the simulation times sampled in this work.

To evaluate the size of the ion clusters, we performed a graph analysis to identify the subsets of connected components, considering ions as nodes in the graph.\cite{tribello_analyzing_2017}
Figure \ref{fig:clusters}~D provides the average largest cluster size as a function of the total number of clusters at each concentration.
This confirms that ion pairs are likely to form in the EDL at 1 M.
When the concentration of ions is increased to 5 M, we observed clusters in the EDL containing around four ions, and the number of these clusters substantially increases when $\sigma<0$.
At~10 M, many clusters are observed in all regions of the solution; however, the largest clusters are typically found in the EDL. 
Moreover, the EDL at the negatively charged graphite surface contains clusters which are significantly larger than the largest clusters observed at the positively charged surface.
A snapshot of one of these extended ion networks is provided inset in Figure \ref{fig:clusters}~B; this highlights the chemical heterogeneity and liquid-like ion connectivity that is typically observed in the assemblies.

\subsection{Water structure at graphite}
Following the evaluation of how surface charge and concentration changes the structural properties of ions at the interface, in this section, we discuss how the interface affects the microscopic water structure in the EDL when compared to the bulk solution.
To this aim, we evaluated variables which are functions of the positions of water O atoms that quantify the relative order of the molecules, as well as the H-bond network in the solvent.
It is useful to compare these analyses of the C$\mu$MD simulations to liquid and solid forms of water.
As such, additional simulations of bulk liquid water, cubic ice (ice $I_c$) and hexagonal ice (ice $I_h$), containing 4,000, 2,744 and 2,880 molecules, respectively, were performed for $5-10$ ns.

\paragraph{Water ordering at the interface}
To quantify the local ordering of water molecules approaching the solid-liquid interface, we computed the approximate two-body excess entropy ($S_2$), adopting the position of oxygen atoms in water as a proxy for the centre of mass of the molecules:\cite{piaggi_enhancing_2017}
\begin{equation}
    S_2 = -2 \pi \rho_{O_w} k_\mathrm{B} \int_0^{r_{lim}} r^2 [g(r) \ln g(r) - g(r) +1] \;\mathrm{d}r
\end{equation}
here, $k_\mathrm{B}$ is Boltzmann's constant and $\rho_{O_w}$ is the atom density in the simulation cell.
$g(r)$ is a radial pair distribution function of the distances, $r$, between $i$ and $j$ pairs of water O atoms:
\begin{equation}
    g(r) = \frac{1}{4 \pi N_{O_w} \rho_{O_w} r^2} \sum_i^{N_{O_w}} \sum_{j \neq i}^{N_{O_w}} \frac{1}{\sqrt{2 \pi} \xi}\exp \left( \frac{-(r-r^{ij})^2}{2 \xi ^2}
    \right)
\end{equation}
Here, $N_{O_w}$ is the total number of water oxygen atoms. We chose $r_{lim}$ to be $0.5$~nm and the broadening parameter, $\xi = 0.015$.
We obtained local averages\cite{lechner_accurate_2008} of $S_2$ according to,
\begin{equation}
    \overline{S_2} = \frac{1}{N_{O_w}}  \sum_i^{N_{O_w}} \left(\frac{S_2^i + \sum_j^{N_{O_w}} f(r^{ij}) S_2^j}{1+ \sum_j^{N_{O_w}} f(r^{ij})} \right)
\end{equation}
Here, $f(r^{ij})$ is a sharp but continuous switching function that identifies water molecules in the first coordination sphere according to,
\begin{equation}
    f_s(r_{ij})=\frac{1-(\frac{r^{ij}}{r^0})^p}{1-(\frac{r^{ij}}{r^0})^{q}}
    \label{eq:switch}
\end{equation}
where $r^0=0.35$~nm, $p=50$ and $q=100$.

\begin{figure}[H]
    \centering
    \includegraphics[width=1\linewidth]{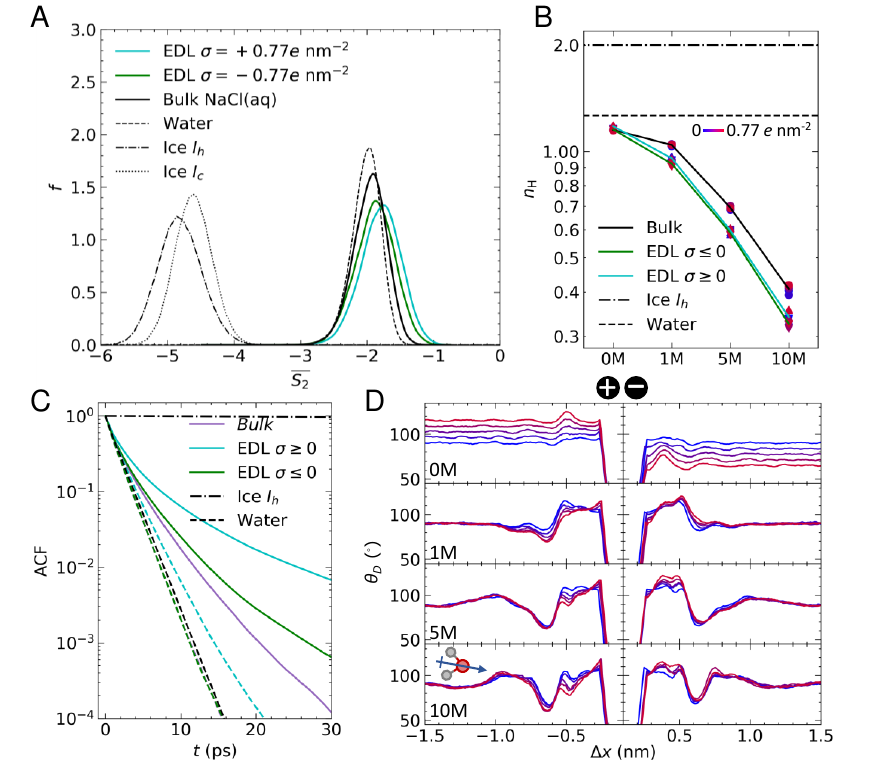}
    \caption{Water structure in the bulk and EDL regions of a C$\mu$MD simulation compared to pure water (Water) and hexagonal ice ($I_h$). A) Water local average approximate pair entropy ($\overline{S_2}$) calculated for 5M NaCl(aq) at graphite with charge density, $|\sigma| = 0.77\,e$ nm$^{-2}$. B) Number of donated hydrogen bonds per water molecule ($n_H$). Colours blue$\rightarrow$red indicate increasing solution charge density as shown by the scale inset. Solid lines show the mean $n_H$ values at each concentration for water in the bulk (circles), negatively (inverted triangles) and positively (triangles) charge interface regions. Uncertainties in the data are on the scale of the size of data points. C) H-bond lifetimes, as indicated by the time-dependent H-bond autocorrelation function (ACF). Data are provided for the case where $|\sigma| = 0.77\,e$ nm$^{-2}$ and ion concentrations are 0M (dashed lines; the curves for bulk and EDL $\sigma \leq 0$ are nearly perfectly overlaid on the graph) and 10 M (solid lines). D) Water molecule orientation as indicated by the average angle between the water molecule dipole moment (shown by the arrow on the water molecule inset) and the normal to the carbon basal surface ($\theta_{D}$). The colours indicate surface charge density (see the scale in B). The data in D were smoothed using a zero$^{\mathrm{th}}$-order Savitsky-Golay filter with a 0.05 nm window size.}
    \label{fig:CVs}
\end{figure}

Figure \ref{fig:CVs}~A provides the approximate excess entropy probability distributions, $f(\overline{S_2})$, when $c$(NaCl) is 5 M at the most extreme values of $\sigma$ ($\pm 0.77 \,e$~nm$^{-2}$).
All liquid water states are clearly separated from the $\overline{S_2}$ range of values calculated for solid water phases, demonstrating that the variable adopted differentiates water molecules with different levels of local order.
The presence of ions leads to a shifting of the median $\overline{S_2}$ to larger values when compared to the case of pure water.
In the context of the variable adopted, this suggests that the water network in solution is less ordered than in liquid water.
In the EDL, the distributions are shifted to even higher values of $\overline{S_2}$, particularly so in the case of the positive electrode, indicating further loss of order when compared to molecules in the bulk solution.

In these analyses, the EDL was taken to be the region above the carbon surface encompassing only the first two solution layers; hence, this is the region of the double layer where the water structure is most perturbed when compared to the bulk.
As shown in Figure \ref{fig:SIOPs}~B, which provides $f(\overline{S_2})$ in slices throughout the entire simulation cell $x$-axis, the local structure of water is only significantly perturbed in the immediate vicinity of the carbon substrate.
Given this observation, it is important to assess how significant the presence of ions and surface charge change the water structure, as opposed to the excluded volume effects associated with the water void space occupied by the graphite slab.

Clearly, Figure \ref{fig:CVs}~A identifies a surface charge effect, but to consider the role that ions play in changing the local water order, we computed additional variables, namely the third-order Steinhardt bond orientational order parameter ($q_3$) \cite{steinhardt_bond-orientational_1983}, as well as its local ($lq_3$) and local average ($\overline{q_3}$) values, for all water molecules comprising the bulk solution. 
For the functional form of $q_3$ and $lq_3$, please refer to the PLUMED documentation. \cite{tribello_plumed_2014}
These variables were previously combined with $\overline{S_2}$ to identify water order in different physical states. \cite{fulford_deepice_2019}
All of the distributions for these variables, provided in Figure \ref{fig:SIOPs}~A, indicate a small deviation from the pure water case, although this is minimal when compared to ice and amorphous water (liquid water crash cooled to 100~K during a 10~ns simulation).
From this analysis, it would appear that the surfaces have a greater effect on the local ordering of water molecule centres than the presence of structure-breaking ions.

\paragraph{Water H-bonds}
The intermolecular structure of water is usually described in terms of the H-bond network.
We, therefore, calculated---using MDAnalysis\cite{gowers_mdanalysis_2016,gowers_multiscale_2015}---the number of donated H-bonds per water molecule, $n_\mathrm{H}$, in the bulk and EDL regions of all simulations using a simple geometric criterion for these bonds of the type, O$_D$---H$\cdots$O$_A$, where $D$ and $A$ subscripts refer to the H-bond donor an acceptor, respectively.
H-bonds were assigned when O$_D$ and O$_A$ were within 0.3~nm and the angle, $\angle$O$_D$HO$_A > 150^\circ$.
Figure \ref{fig:SIHbonds} provides the distributions for H-bond distances and angles in the EDL and bulk regions of C$\mu$MD simulations, as well as for ice $I_h$ and pure liquid water.

Figure \ref{fig:CVs}~B shows that irrespective of the applied surface charge, the mean $n_\mathrm{H}$ for water in the bulk and EDL regions of C$\mu$MD simulations in the absence of ions are close to the values in pure liquid water.
The small deviation from the homogeneous liquid case is likely to result from the excluded volume effects associated with the selection of water molecules in regions of the simulation cell $x$-axis that creates (artificial) excluded volumes, even in the case of what we describe as the \textit{bulk}.
Slightly more than one H-bond per water molecule is observed, which is understandably lower than the expected value of two for ice $I_h$.

In contrast to the structural variables discussed above, $n_\mathrm{H}$ is far more sensitive to the solution concentration and less sensitive to the magnitude and sign of $\sigma$.
At 1~M, we find a difference between $n_\mathrm{H}$ in the EDL when compared to the bulk by around $n_\mathrm{H}=0.1$.
This difference was approximately the same at all levels of concentration; however, $n_\mathrm{H}$ in all regions of the simulations decreased as the concentration of ions increased.
At 10~M, $n_\mathrm{H}$ is approximately $0.35-0.4$, suggesting a near complete breaking of the H-bond structure at the highest concentrations.
The presence of ions and their assemblies, as well as associated local electric fields, greatly perturbs the water structure from the pure solvent, which ultimately has implications for these systems and their performance as conductors of electrical charge.

The lifetime for H-bonds was determined according to the autocorrelation function (ACF):
\begin{equation}
    ACF(\tau) = \left< \frac{H_{ij}(t_0) H_{ij}(t_0 + \tau)}{H_{ij}(t_0)^2} \right>
\end{equation}
where $\tau$ is a time lag in the data, $t_0$ indicates a time origin and $H_{ij}$ signifies the presence of an assigned H-bond between molecules $i$ and $j$, taking a binary value of zero or one according to the H-bond distance and angle cut-offs described above.
The H-bond lifetimes are provided in Figure \ref{fig:CVs}~C for pure liquid water, ice and the bulk and EDL regions of simulations at 0 and 10 M with $\sigma = \pm 0.77 \,e$~nm$^{-2}$.

In the absence of ions, the H-bond lifetime for water in the bulk and at the negative electrode is identical to the lifetime of H-bonds in pure water.
At the positive electrode, however, the H-bond lifetime is extended upon the application of a large surface charge.
This is most likely due to the increased binding strength of water O atoms to the charged graphite surface.
At 10 M, we find that the H-bond lifetimes in the bulk are extended cf. 0 M, and there is a divergence in the lifetimes in the bulk and at the negative electrode.
These lifetimes, however, are still lower than the average H-bond lifetime at the positive electrode, which can extend to $\sim 10^2$~ps.
Simulations indicate that the orientation of water molecules at charged surfaces determines the propensity for heterogeneous ice crystallisation, with positively charged silver iodide surfaces suggested to promote ice nucleation. \cite{glatz_surface_2016} 
A stabilisation of the H-bond lifetime at positively charged surfaces potentially has additional ramifications on the ability of these charged substrates to promote ice nucleation. 
In light of our findings, it would be useful to test how these phenomena \textit{and} the presence of ions in solution control ice crystallisation rates.

\paragraph{Water dipole moments}
We characterised the orientation of water molecules with respect to the graphite surface plane by computing the angle between water molecule dipole moments and the normal to the surface, $\theta_D$.
The calculation was implemented such that if the dipole moment was perfectly perpendicular to the surface normal and pointing towards the surface (see the water dipole moment arrow inset of Figure \ref{fig:CVs}~D), the value of $\theta_D$ is zero and $180^\circ$ if the dipole moment vector points away from the surface.
$\theta_D=90^\circ$, therefore, indicates water molecules with dipole moments that are, on average, aligned parallel with the basal surface\RV{, and/or indicates no preference for the orientation of water molecules at the surface.}

Figure \ref{fig:CVs}~D provides the mean $\theta_D$ as a function of $\Delta x$.
Before discussing the effect of surface charge and solution concentration on these results, it is useful to discuss an important feature of the curves at 0 M.
In this subset of simulations, no ions are present, and we did not adopt C$\mu$MD to study the effect of charged surfaces on the water structure. As such, there is no ionic reservoir in this system.
When equal but opposite signs of surface charge are applied to opposite faces of the graphite slab, electric fields are induced that span the periodic boundaries in $x$.
This effect is evident in the $\theta_D$ curves at 0 M, where the neutral graphite case (see the symmetric blue curves at both surfaces) indicates that $\theta_D=90^\circ$ in the bulk, but where positive and negative deviations in the mean $\theta_D$ are found at the positive and negative electrode, respectively.
It is possible to avoid these electrical artefacts by removing the periodic boundaries in $x$ and/or by including artificial electrical insulating layers parallel to the graphite surface. 
For the purposes of this study, this was not necessary, and, importantly, these effects are removed by the presence of the ionic reservoir.
In terms of the discussion of $\theta_D$ at 0 M, we compare the features of the distributions in the EDL with respect to the values in the bulk in order to understand how the surface controls the orientation of water; furthermore, we do not believe that the dipole associated with the graphite slab has a significant effect on the analysis of EDL solution thermodynamics, discussed in the following section.

The $\theta_D$ distributions indicate that the surface, in the absence of ions, leads to no significant ordering of water molecule dipole moments with respect to the surface plane, as expected with the adopted force field\cite{wu_graphitic_2013} and shown from simulations elsewhere. \cite{elliott_qmmd_2020}
Furthermore, negative surface charges give rise to increased $\theta_D$ in the first solution layer(s), confirming that water molecules tend to point away from the surface (compared to the bulk) in order to maximise the interactions between H atoms and the surface.
At the positive electrode, a maximum occurs at $\Delta x \approx -0.5$~nm; here, there is a depletion in the water density (see Figure \ref{fig:d0M}), which perhaps allows water to restructure more freely to screen the surface potential.

In the presence of ions, $\theta_D$ angles are obtuse in the first two solution layers, regardless of the sign of the surface charge; however, maximum values are observed on the negative electrode.
In all curves, a minimum occurs at $\Delta x \approx -0.65$~nm, which is where the second peak in water density profiles is observed when moving away from the surface.
An acute $\theta_D$ was also observed for 1 M solutions in contact with graphene using quantum mechanical MD simulations. \cite{elliott_qmmd_2020}
This result indicates that the mean orientation of water dipoles in the first and second solution layers differs by $40-50^\circ$.
The complex EDL structure is evident in the $\theta_D$ curves, notably at 10 M, where large fluctuations in the angle distributions are evident within 0.5~nm from graphite.
Finally, very little ordering of water is observed beyond $\sim 2$~nm from the substrate at the highest concentrations.

\subsection{Solution thermodynamics}
As well as the capability to undergo rapid charge/discharge cycling, carbon-electrolyte interfaces can be exploited for applications to promote chemical reactions. \cite{tian_nanoengineering_2019,julkapli_graphene_2015}
Understanding the thermodynamic properties of the interface is essential in this regard.
In the following section, we characterise the electrochemical properties of the interface, with a focus on the solution side of the EDL.

\begin{figure}[H]
    \centering
    \includegraphics[width=0.85\linewidth]{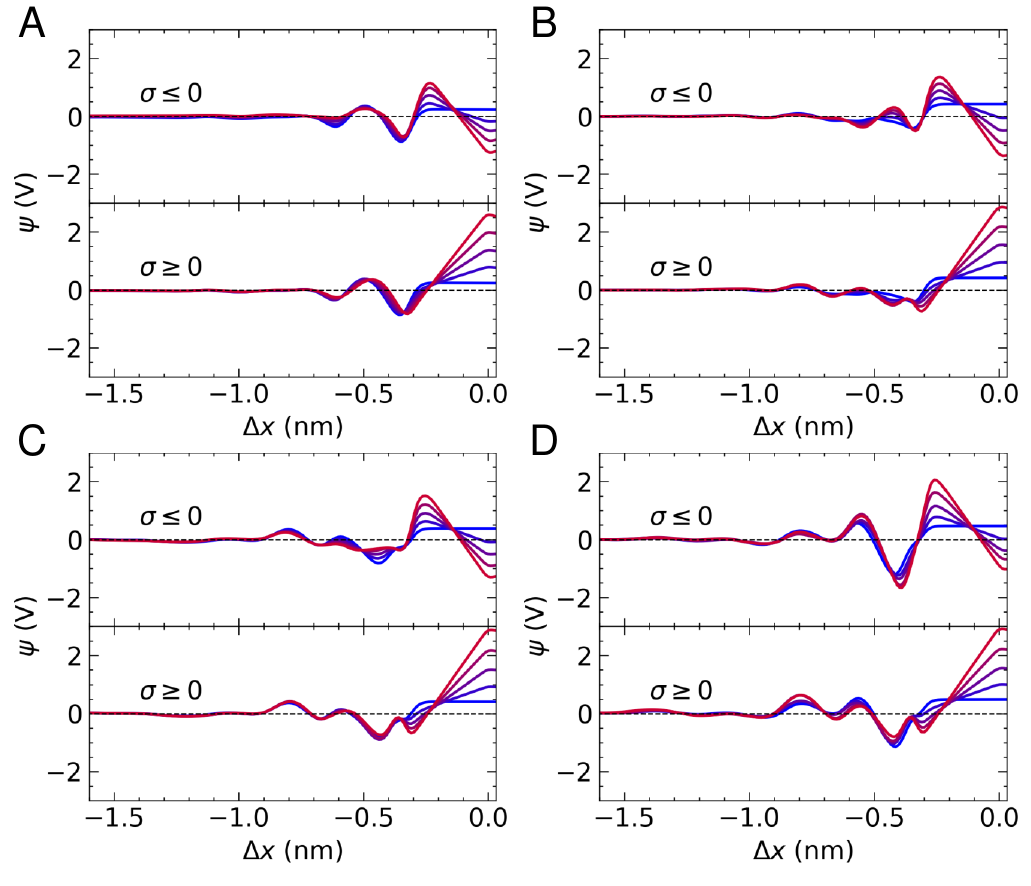}
    \caption{Interface electric potential as a function of distance from the electrode, $\Delta x$. Colours blue$\rightarrow$red indicate increasing solution charge densities from $|\sigma | =0 \rightarrow 077~$e~nm$^{-2}$. A, B, C and D are taken from simulations where the target bulk ion concentration was 0, 1, 5 and 10~M, respectively. }
    \label{fig:potential}
\end{figure}

\paragraph{Electric potential at the interface}
The capacity for the interface to store charge, $C=\sigma/\Delta \psi^0$, where $\Delta \psi^0$ is the electric potential change (usually termed the `potential drop') across the interface with an applied surface charge minus the potential drop in the absence of a surface charge.
Poisson's equation relates the electric potential to the charge density ($\rho_q$) according to,
\begin{equation}
    \frac{\mathrm{d}^2\psi(x)}{\mathrm{d}x^2} = -\frac{\mathrm{d}E(x)}{\mathrm{d}x} = -\frac{\rho_q(x)}{\varepsilon}
    \label{eqn:poisson}
\end{equation}
where $E$ is the electric field and $\varepsilon$ is the permittivity of the medium.
Here we take $\varepsilon = \varepsilon_0$, the permittivity of free space, because the full solution charge density is used in the calculation.
It is important to recognise that this analysis provides the capacitance associated with the solution side of the EDL in contact with a uniformly charged surface.
An additional contribution to the total capacitance comes from the density of electron states in the substrate, which represents a minor contribution to the interfacial capacitance under the conditions studied. \cite{pasquale_constant_2022,finney_bridging_2022}

Figure \ref{fig:potential} provides $\psi(x)$ for all systems and all applied charges.
In the case of pure water, we find that the potential drop across the interface, defined as $\Delta \psi = \psi(\Delta x=0)-\psi^b$ (where $\psi^b$ is the electrostatic potential in the bulk), in the absence of surface charge is $0.23$~V.
As the surface charge is increased to $\pm 0.77 \,e$~nm$^{-2}$, $\Delta \psi$ was calculated as $-1.2$~V and $2.6$~V on the negative and positive electrode, respectively.
Two minima are observed in Figure \ref{fig:potential}~A, in accordance with the maxima in the water density profiles (see Figure \ref{fig:densities}).
The application of surface charge perturbs these densities, as discussed above, which changes the position of the first minima in $\psi(x)$ (from the surface) and makes the second minimum shallow when $\sigma< 0$.

The presence of ions induces additional fluctuations in $\psi(x)$ when compared to the pure water case. 
These extend to around 1.5~nm from the carbon surface at the highest concentrations, although the amplitude of the fluctuations is not particularly correlated with concentration, due to the fact that local electric fields are determined by the \textit{total} solution charge density.
Some notable features of the $\psi(x)$ curves are the fact that cation adsorption in the absence of surface charge increases $\Delta \psi$ to $\sim 0.42$ ~V, in good agreement with studies of aqueous solutions at carbon surfaces using different models for the interface. \cite{elliott_electrochemical_2022}
In addition, a minimum emerges at $\Delta x \approx -0.3$~nm at the positive electrode, associated with a depletion of cations and accumulation of anions.
Furthermore, a deep minimum is observed at 10 M when $\Delta x \approx -0.4$~nm on the negative electrode, which can be attributed to the increasing anion concentration and water restructuring that occurs at this surface.

When $\sigma=+0.77\,e$~nm$^{-2}$, $\Delta \psi$ at the positive electrode ranges from $2.85-2.9$~V across the range of concentrations investigated.
Using the expression for capacitance reported above, this equates to a solution-side capacitance of $\sim5$ \textmu F~cm$^{-2}$.
$\Delta \psi$ at the negatively charged electrode, instead, ranges from $-0.92$ to $-1.34$~V, corresponding to a capacitance of $C=7-9$ \textmu F~cm$^{-2}$, indicating that graphite has a greater capacity to store charge when negative charges are applied. $C$ evaluated in these simulations is consistent with estimates from experiments. \cite{finney_electrochemistry_2021}
Moreover, the increased capacity to store ionic charge at the negative electrode is consistent with results elsewhere. \cite{pasquale_constant_2022,dockal_molecular_2019}
A result worthy of note is that the concentration of ions has very little effect on the ability of graphite to store charge over the range of molarities considered, which was also found for graphene. \cite{pasquale_constant_2022}
Indeed, increasing the charge capacity of the carbon-solution interface typically requires tuning the properties of the solute and the substrate to increase the overall capacitance of the system, rather than simply changing the concentration of the solution. \cite{ji_capacitance_2014,fedorov_ionic_2014}

\paragraph{Ion activity coefficients}

\begin{figure}[H]
    \centering
    \includegraphics[width=0.95\linewidth]{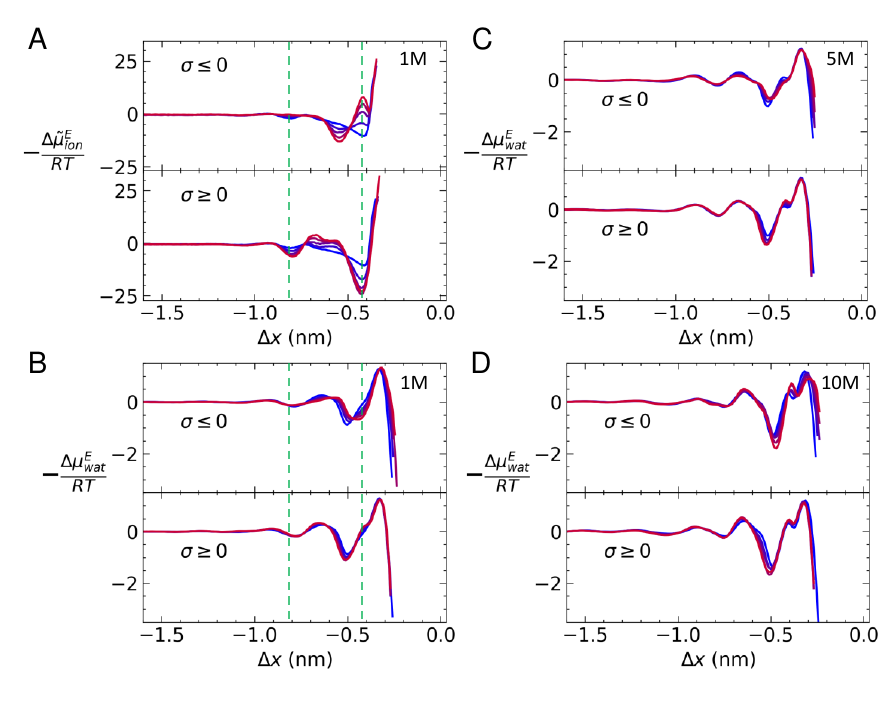}
    \caption{A) The excess ion electrochemical potential $-\Delta \tilde{\mu}_{ion}^E/RT = 2\ln (\gamma_{ion})$ in solution (1M) as a function of distance from the graphite basal surface ($\Delta x$). B--C) The excess water chemical potential $-\Delta \mu_{wat}^E/RT = \ln (\gamma(x)/\gamma^b)$ for water molecules as a function of $\Delta x$ when the target bulk solution concentration was 1, 5 and 10 M, respectively. In all plots, the top and bottom panels provide data for solutions at negatively and positively charged surfaces, respectively; blue$\rightarrow$red colours indicate $| \sigma | = 0-0.77\,e$ nm$^{-2}$. Green dashed lines highlight the position of the first two minima in $-\Delta \mu_{ion}^E/RT$ when $\sigma=0$.}
    \label{fig:activities}
\end{figure}

The chemical potential of species $i$, $\mu_i$, is defined as the change in free energy associated with a variation in the number of $i$ molecules and represents the ability of that species to undergo a physical-chemical transformation. 
In the presence of an electric field, when $i$ is a charged species, the same information is captured by the electrochemical potential, which accounts for the additional energetic contributions to insert/remove a charged particle to/from the system:
\begin{align}
\begin{split}
    \tilde{\mu}_i &= \mu_i^0 + RT \ln a_i + z_iF \psi\\
    &= \mu_i^0 + RT \ln m_i +  RT \ln \gamma_i + z_iF \psi    
\end{split}
\label{eqn:ecp}
\end{align}
In the above equation, $\mu^0$ is a reference chemical potential.
The second term on the right of the equation provides the energy associated with particle exchange in non-ideal solutions, where $R$, $T$ and $a$ indicate the gas constant, temperature and solute activity, respectively.
This term can be expanded to account for the \textit{ideal} and \textit{excess} chemical potential, which are functions of the total concentration, in this case, the solution molality, $m$ (strictly, this is a unitless quantity defining the mole fraction of solute in solution compared to the standard state of 1 mol/kg), and the activity coefficient, $\gamma$.
The final term defines the work to transfer a particle with charge $z$ into the system with electrostatic potential, $\psi$. Faraday's constant, $F$, ensures that the term has the correct energy units.

In order to determine $\tilde{\mu_i}$ for ions in our simulations, an activity model is required.
Zimmerman et al. provided an analytical formula to calculate $\mu$ for ions in NaCl(aq) as a function of ion molality, $m_{ion}$, by fitting to simulation data: \cite{moucka_molecular_2013,mester_mean_2015,zimmermann_nacl_2018}
\begin{equation}
    \mu_{ion} = \mu_{ion}^0 + 2RT \ln m_{ion} +  2RT \ln \gamma_{ion}
    \label{eqn:activity}
\end{equation}
where,
\begin{equation}
    \log_{10} \left( \gamma_{ion} \right) = \frac{a \sqrt{m}}{1+b \sqrt{m}} + cm
\end{equation}
In these equations, $\mu^0_{ion} = -391.6$ kJ mol$^{-1}$,\cite{mester_mean_2015} $a=0.568$ mol$^{-1/2}$ kg$^{1/2}$, $b=1.17769$ mol$^{1/2}$ kg$^{-1/2}$ and $c=0.177157$ mol$^{-1}$ kg.
It is important to recognise that this activity model assumes changes to the solution density and dielectric constant are only a function of the solution composition.
This model can be extended\cite{finney_electrochemistry_2021} to account for the effect of electric fields and associated varying ion molalities that occur on approach to the graphite surface according to,
\begin{align}
\begin{split}
    \tilde{\mu}_{ion}(x) &= \mu_{ion}^0 + RT \ln m_{Na}(x) +  RT \ln \gamma_{ion}(m_{Na}(x)) + \omega F \psi(x) \\
    &+  RT \ln m_{Cl}(x) +  RT \ln \gamma_{ion}(m_{Cl}(x)) - (1- \omega)F \psi(x)
\end{split}
\label{eqn:ecp_edl}
\end{align}
where subscript labels indicate Na$^+$ or Cl$^-$ molalities and $\omega(x) = m_{Na}(x)/(m_{Na}(x)+m_{Cl}(x))$.
In the limiting case where the molalities of cations and anions are equal locally, Equation \ref{eqn:ecp_edl} reduces to Equation \ref{eqn:activity}.

For the case of 1 M NaCl(aq), we determined $\tilde{\mu}_{ion}(x) \approx -392$ kJ mol$^{-1}$ in the bulk where $\psi(x)=0$, in good agreement with the expected chemical potential from the model by Zimmerman et al.\cite{zimmermann_nacl_2018} for homogeneous solutions with $m_{ion} \approx 1.3$ mol kg$^{-1}$.
$\gamma_{ion}=0.9$ under these conditions, which we approximate to a value of one for the subsequent analyses, such that $\tilde{\mu}^b_{ion} \approx \mu_{ion}^0$, with $\tilde{\mu}^b_{ion}$ representing the electrochemical potential of ions in the bulk. The energy change associated with $2RT \ln \gamma_{ion}$ when $\gamma_{ion}=0.9$ is $\sim 0.5$ kJ mol$^{-1}$.

In our simulations, the chemical potential of ions and water as a function of $x$ in the steady state is constant.\cite{karmakar_non-equilibrium_2023}
Therefore, with knowledge of the ion molalities and electric potential in the EDL, Equation \ref{eqn:ecp_edl} can be rearranged to determine how the presence of the surface---where the density of ions and dielectric constant of the solution are changing compared with the bulk---affects the activity of ions as captured by $\gamma_{ion}$:
\begin{align}
\begin{split}
    -RT \ln \left[ \gamma_{ion}(m_{Na}(x)) \gamma_{ion}(m_{Cl}(x) ) \right] = RT \left[ \ln m_{Na}(x) m_{Cl}(x) \right]+ (2\omega -1)F \psi(x) 
\end{split}
\label{eqn:ecp_equal}
\end{align}
We label this quantity $\Delta \tilde{\mu}^E_{ion}$.
Figure \ref{fig:activities}~A provides $-\Delta \tilde{\mu}^E_{ion}/RT= 2 \ln(\gamma_{ion})$ at graphite with varying charge density when $c(\mathrm{NaCl})=1$~M. 
On approach to the surface, there is a small minimum around $\Delta x = -0.8$ nm when $\sigma=0$, which is consistent with the position of a second cation-rich solution layer above the surface.
This is followed by a gradual decrease in $2 \ln(\gamma_{ion})$ towards a second minimum around $\Delta x = -0.4$ nm.
This minimum resides between the maxima for the densities of the first cation- and anion-rich solution layers in the EDL (see Figure \ref{fig:d1M}).

When a positive charge is applied to graphite, as shown in the bottom panel of Figure \ref{fig:activities}~A, $2 \ln(\gamma_{ion})$ becomes more negative to around $-25RT$; this is due to the positive surface charge pushing and pulling Na$^+$ and Cl$^-$ away from and towards the surface (see Figure \ref{fig:densdiff}).
In addition, there is a small increase in $2 \ln(\gamma_{ion})$ at $\Delta x = -0.65$ nm, due to a relatively high mole fraction of Na$^+$ in this region.
On the negatively charged surface, the applied potential displaces Cl$^-$, which leads to a decrease in the anion density at $\Delta x = -0.4$ nm and an increase at $\Delta x = -0.55$ nm compared to the case when $\sigma=0$ (see Figure \ref{fig:densdiff}); this results in positive and negative increases to $2 \ln(\gamma_{ion})$, respectively.

Figure \ref{fig:cp-contributions} provides the contributions to  $\Delta \tilde{\mu}_{ion}(x)$ for a single case where $c=1$ M and $\sigma=+0.77\,e$ nm$^{-2}$.
This shows that it is essential to account for the effect of the surface excluded volume and charge on the structure of the solution when determining the activities of ions.
In particular, in the presence of an applied surface potential, the contribution to $\Delta \tilde{\mu}_{ion}(x)$ from the electric potential drop can be as significant as the changing mole fraction of solute at the surface.

\paragraph{Water activity coefficients}
We now consider how charged surfaces affect water activity coefficients in the EDL when compared to the bulk. 
We note that, for an electrically neutral species such as water, the electrochemical potential reduces to the chemical potential even in the presence of a charged surface.
To determine the chemical potential of SPC/E water in NaCl(aq) without an accurate activity model, we make use of the fact that in our C$\mu$MD simulations, the chemical potential for water molecules in the EDL and bulk are equal.
We can estimate the potential of mean force ($\mathcal{W}_{wat}$) to transfer water molecules from the bulk to the EDL according to,
\begin{equation}
    \Delta \mathcal{W}_{wat}(x) = -RT \ln \left( \frac{p_{wat}(x)}{p_{wat}^b} \right)
    \label{eqn:pmf}
\end{equation}
where $p_{wat}$ represents the probability density of observing water molecules at position $x$, and the superscript $b$ indicates the probability density at a point in $x$ representative of the bulk solution.
As such, $\Delta \mathcal{W}_{wat}$ provides a proxy for $\Delta A$: the Helmholtz free energy change for the transformation under question.

Given that $\Delta A = n \Delta \mu_{wat} = -RT \ln K$, where $n$ is the number of moles of water and $K=a_{wat}(x)/a_{wat}^b$ (where $a$ indicates activity) which is the equilibrium constant for the transfer of one water molecule from position $x$ to the solution bulk, we can also write,
\begin{align}
\begin{split}
    \Delta \mathcal{W}_{wat}(x) &= -RT \ln  \left( \frac{\chi_{wat} (x)}{\chi_{wat} ^b} \right) -RT \ln  \left( \frac{\gamma_{wat}(x)}{\gamma_{wat}^b} \right)   \\
    &= \Delta \mu_{wat}^I + \Delta \mu_{wat}^E
\end{split}
\label{eqn:fewat}
\end{align}
where $\chi_{wat}$ and $\gamma_{wat}$ are the mole fraction and activity coefficient for water molecules in solution, respectively. 
Hence, by combining equations \ref{eqn:pmf} and \ref{eqn:fewat} we can evaluate $\Delta \mu_{wat}^E$, which indicates how $\gamma_{wat}$ changes in comparison to $\gamma_{wat}^b$.

Figure \ref{fig:activities}~B--D provides $-\Delta \mu_{wat}^E(x)/RT$ for systems where the ion concentration varies from 1 to 10 M.
At 1 M, two minima are observed in $\ln(\gamma_{wat}/\gamma^b_{wat})$ at $\Delta x = -0.5$ and $-0.75$~nm that are within $RT$ of the bulk value.
The contributions to $\Delta \mathcal{W}(x)$ in the case where $c=1$ M and $\sigma=+0.77\,e$ nm$^{-2}$ are provided in Figure \ref{fig:cp-wat-contributions}, which indicate that the first minimum from the surface arises due to a relatively high value of $\Delta \mu_{wat}^I(x)$ in this region, associated with a decreased water mole fraction, as well as effects associated with the electric fields close to the carbon basal plane.
The minimum at $\Delta x = -0.75$ nm, however, occurs in a region where the water mole fraction is not significantly different from the bulk value and is due to the structuring of ions and electric fields locally.
As the magnitude of the applied charge increases, small shifts to the position of the minima occur, associated with changes to the water density.
The maximum in $\ln(\gamma_{wat}/\gamma^b_{wat})$ occurs around $\Delta x =- 0.3$ nm; this conforms to the minimum in $\Delta \mathcal{W}$, and represents the first water layer adsorbed at the graphite surface (see Figures \ref{fig:d1M} \ref{fig:cp-wat-contributions}).
This indicates that $\gamma_{wat}$ is $3-4$ greater in the innermost EDL solution layer than in the bulk.

As the concentration of ions is increased, the positions of the maxima and minima in $-\Delta \mu_{wat}^E(x)/RT$ are unchanged. 
Additional fluctuations beyond $\Delta x \approx -1$ nm are observed at the highest concentrations, which are only partly associated with changes to $\Delta \mu_{wat}^I(x)$ (see Figure \ref{fig:cp-wat-contributions-10M}). 
At 10 M, the most negative minimum in Figure \ref{fig:activities}~D suggests that $\gamma_{wat}/\gamma^b_{wat}=0.2$.
It is also apparent, at the highest concentration, that an additional minimum in $-\Delta \mu_{wat}^E(x)/RT$ occurs around $\Delta x = -0.35$ nm.
This can be attributed to an increase in $\Delta \mu_{wat}^I$ when compared with lower concentrations, concomitant with a decrease in $\chi_{wat}$ compared with $\chi^b_{wat}$.
Changes to the local structure of the solution on increasing surface charge density tend to be greatest at the highest bulk concentrations (see Figure \ref{fig:densdiff}), so it is perhaps not surprising that a richer behaviour in the $\Delta \mu_{wat}^E$ curves emerges at 10 M as a function of $\sigma$.


\section{Conclusions}
Understanding the properties of the carbon-electrolyte interface is important for a range of applications of these systems to facilitate, e.g., charge storage and chemical reactions.
The C$\mu$MD simulations we have performed in this work provide an atomic scale resolution of the interface of graphite with NaCl(aq) where the concentration of ions and surface charge was varied.
The simulations allow us to investigate how the asymmetric but cooperative adsorption of ions in the EDL, under the effect of an applied potential, affects the structure, dynamics and thermodynamic properties of the interface at a constant thermodynamic driving force for adsorption (defined by the chemical potential of ions in the bulk solution).

Our simulations indicate that increasing the magnitude of the surface charge is analogous to increasing the concentration of ions in solution; both changes give rise to a complex, multi-layered solution structure comprising cation- and anion-rich solution layers due to the finite size of ions and the partial saturation of solution layers with ions, that is not readily captured by mean-field models of the EDL.
Perturbations to the solution structure typically extend $1-2$~nm from the surface.
Interestingly, the presence of a relatively low concentration of ions decreases the intensity of fluctuations in water densities in the EDL compared to the case where no ions are present.

Liquid-like NaCl clusters have been observed in bulk NaCl(aq) solutions at relatively high concentrations \cite{hwang_hydration_2021,finney_multiple_2022}, and our previous work on graphite identified that these clusters are stabilised in the EDL.\cite{finney_electrochemistry_2021}
In the present study, we demonstrated that negative surface charges increase the number and size of these networks, which can include tens of ions at moderate supersaturations.
Furthermore, at negative electrodes, the local ion density in these networks increases, as indicated by changes to the average cation-anion coordination number.
This result raises important questions regarding the ability of charged surfaces to induce NaCl crystallisation.

Our analyses of water structure in the EDL and the lifetime of H-bonds indicate that positive electrodes can induce the reorientation of water molecules at the surface and increase the lifetime of H-bonded networks in this region.
These effects, in turn, have potential implications for the crystallisation of ice in the presence of charged carbon substrates, as well as for the role that these interfaces play in catalysis.
\RV{It is important to note that the water model we adopt is constrained to its equilibrium, bulk liquid water geometry and the partial charges on O and H atoms are fixed.
Future studies should consider how constrained geometry, non-polarisable water models affect the trends found in this work, and how different models for carbon-water interactions affect the thermodynamic properties evaluated here.}

Although our analysis of the electrical properties of the interface indicates a small increase in the capacity of the negative electrode to store charge, the difference in capacitance was $2-4$ \textmu F cm$^{-2}$, with solution concentration playing only a small role in increasing the ability for the negative electrode to accumulate ions.
This is perhaps not surprising because, at molar concentrations, the affinity of graphite for cations and the cooperative adsorption of anions lends the first solution layers already partially saturated with ions in the absence of surface charge.
Analysis of how water and ion activity coefficients deviate from the bulk values, when the electric potential in the EDL is accounted for, indicates how the excess chemical potential for water decreases in the first solution layers adjacent to the surface, concomitant with an increase of the water density in this region, as well as how the changing ion densities induces fluctuations in water activity ratios in the EDL as the concentration of ions increases.

In summary, the complex interplay of solution concentration and surface charge effects provides a picture of the EDL that is difficult to obtain in experiments and from mean-field models.
We hope that the questions raised in this work provide inspiration for further simulation and experimental studies of this system.
\label{sec:conclusions}


\begin{acknowledgement}

The authors acknowledge funding from the Crystallisation in the Real World EPSRC Programme Grant (Grant EP/R018820/1) and the ht-MATTER UKRI Frontier Research Guarantee Grant (EP/X033139/1). The authors acknowledge the use of the UCL Myriad High Throughput Computing Facility (Myriad@UCL), and associated support services, in the completion of this work.

\end{acknowledgement}

\begin{suppinfo}
Additional figures are included in the associated supporting information.

\end{suppinfo}

\section{Data Availability}
GROMACS input and example output files, including the force field parameters necessary to reproduce the simulation results reported in this paper, are available on github (see https://github.com/aaronrfinney/CmuMD-NaCl\_at\_graphite).
The PLUMED input files are also accessible via PLUMED-NEST (www.plumed-nest.org \cite{the_plumed_consortium_promoting_2019}), the public repository for the PLUMED consortium, using the project ID, \RV{plumID:23.027}. 
Details on how to use and implement the C$\mu$MD method within PLUMED is available on github \\ (see https://github.com/mme-ucl/CmuMD). 
\RV{
\section{Author Contributions}
A.R.F. and M.S. designed the research. A.R.F. performed the research and analyses. A.R.F. and M.S. wrote and edited the paper.

\section{Conflict of Interest}
The authors declare no conflict of interest.
}

\bibliography{ref}

\clearpage
\pagenumbering{arabic}
\appendix
\setcounter{figure}{0} 
\setcounter{table}{0} 
\setcounter{equation}{0} 
\renewcommand\thefigure{S\arabic{figure}}
\renewcommand\thetable{S\arabic{table}}

\newpage
\onecolumn
{
\centering
~~
\section{Properties of aqueous electrolyte solutions at carbon electrodes: effects of concentration and surface charge on solution structure, ion clustering and thermodynamics in the electric double layer}

\vspace{0.5 cm}
{\Large \underline{\underline{Supporting Information}}} 
\vspace{1 cm}

{\large Aaron R. Finney and Matteo Salvalaglio}
\vspace{0.5 cm}

\textit{Thomas Young Centre and Department of Chemical Engineering, University College London, London WC1E~7JE, United Kingdom}
\vspace{0.5 cm}

E-mail: a.finney@ucl.ac.uk; m.salvalaglio@ucl.ac.uk

}
\vspace{1 cm}
\etocdepthtag.toc{mtappendix}
\etocsettagdepth{mtchapter}{none}
\etocsettagdepth{mtappendix}{paragraph}
\etocsetnexttocdepth {paragraph}

~~

\section{Additional Figures}
\addcontentsline{toc}{section}{Additional Figures}

\begin{figure}[H]
    \centering
    \includegraphics[width=1\linewidth]{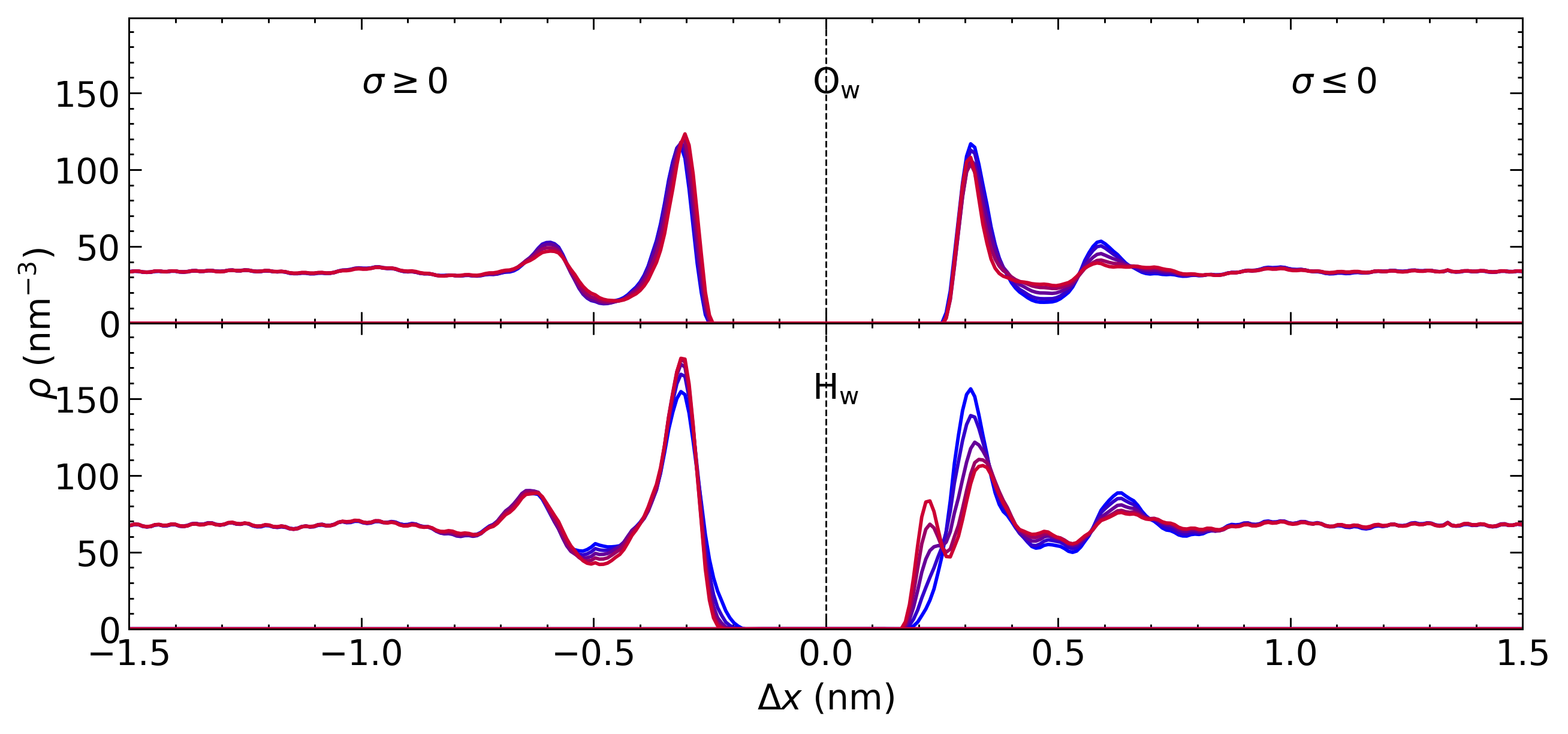}
    \caption{Water atom densities, $\rho$, in the steady state determined as a function of distance from the graphite electrode, $\Delta x$. The atom type is indicated on each panel. Colours blue$\rightarrow$red indicate increasing graphite surface charge densities from $|\sigma | =0 \rightarrow 0.77\; e$~nm$^{-2}$, with data on the left and right of each panel pertaining to simulations with positive and negative applied surface charges, respectively.}
    \label{fig:d0M}
\end{figure}

\begin{figure}[H]
    \centering
    \includegraphics[width=1\linewidth]{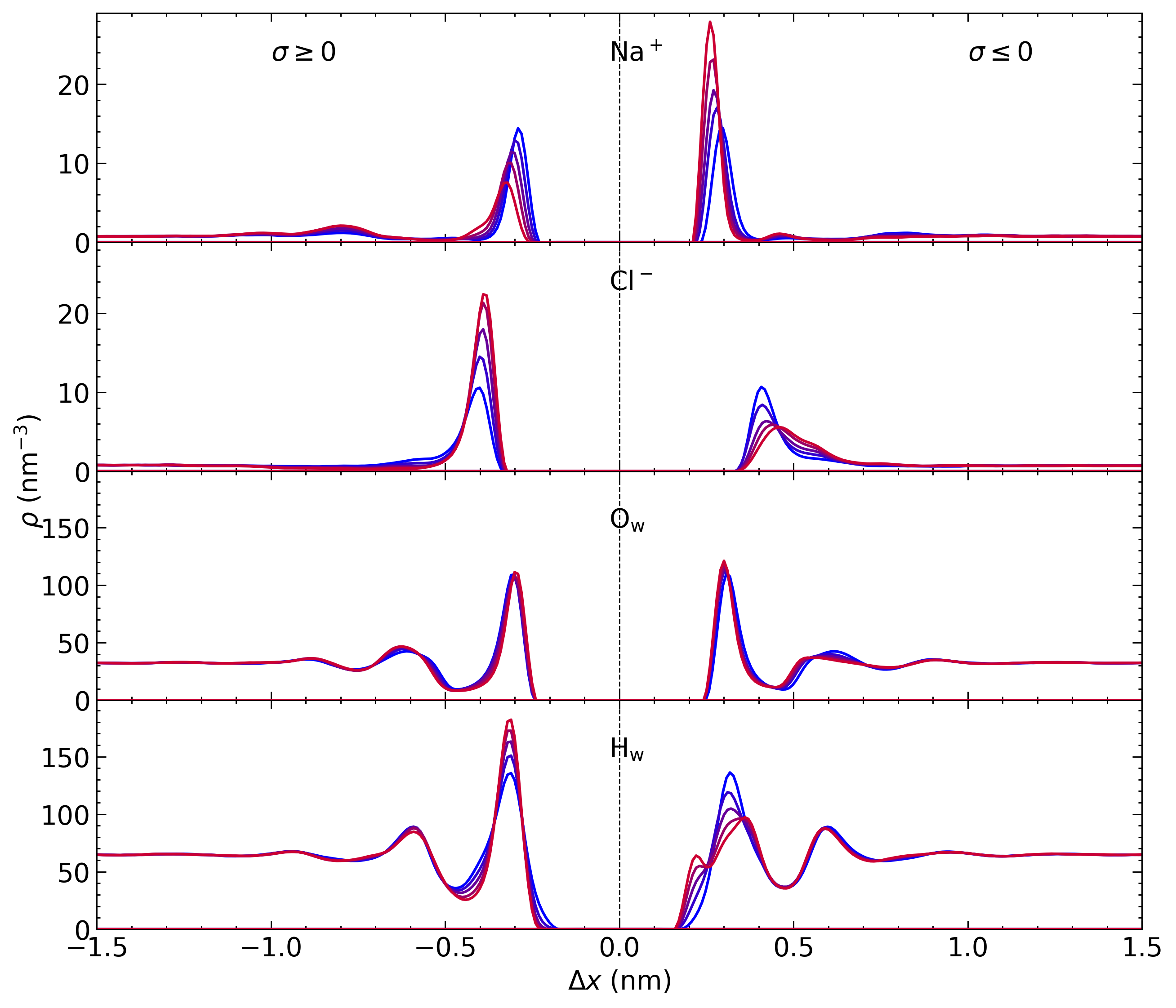}
    \caption{Water atom and ion solution densities, $\rho$, in the steady state determined as a function of distance from the graphite electrode, $\Delta x$, where the target bulk ion concentration was 1 M. The atom type is indicated on each panel. Colours blue$\rightarrow$red indicate increasing graphite surface charge densities from $|\sigma | =0 \rightarrow 0.77\; e$~nm$^{-2}$, with data on the left and right of each panel pertaining to simulations with positive and negative applied surface charges, respectively.}
    \label{fig:d1M}
\end{figure}

\begin{figure}[H]
    \centering
    \includegraphics[width=1\linewidth]{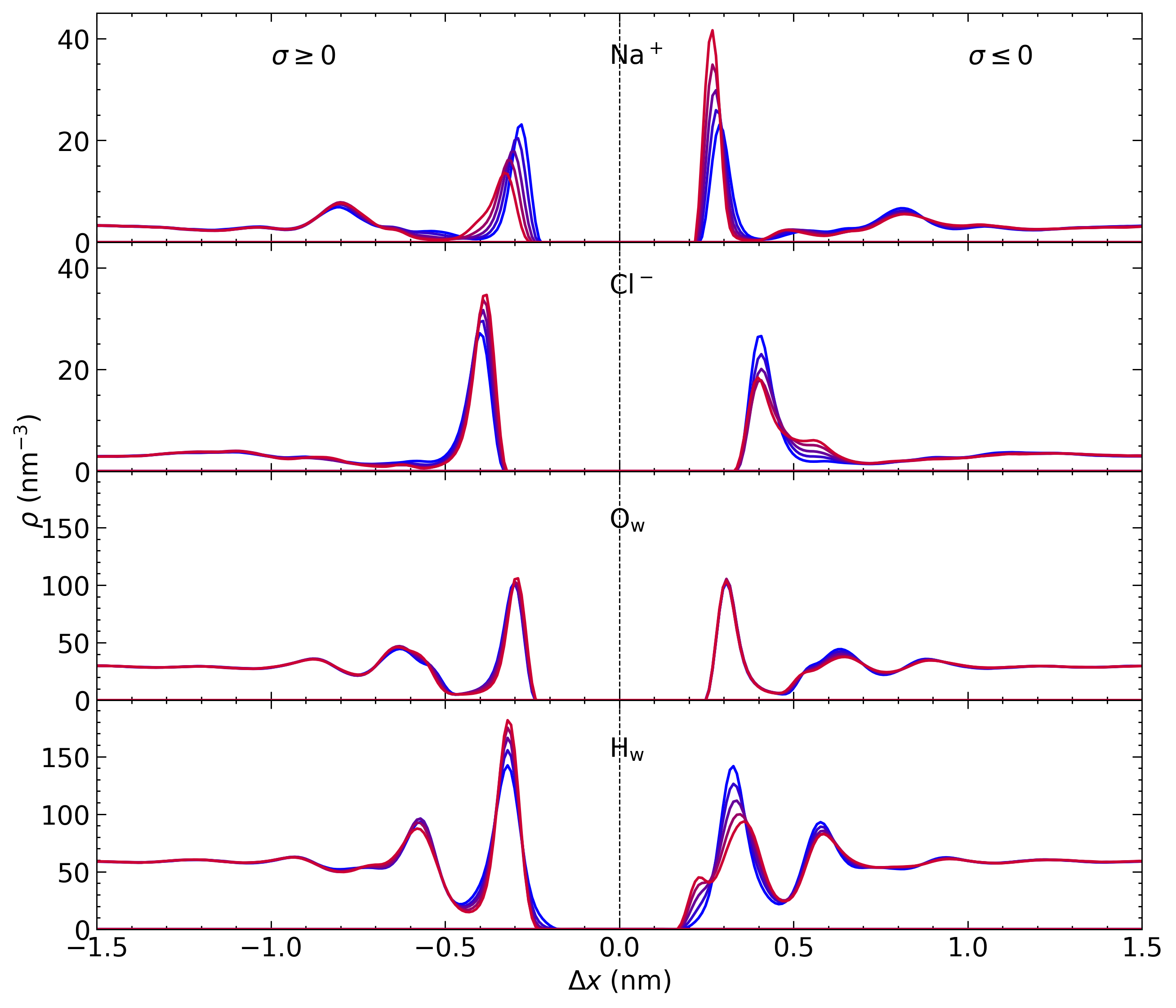}
    \caption{Water atom and ion solution densities, $\rho$, in the steady state determined as a function of distance from the graphite electrode, $\Delta x$, where the target bulk ion concentration was 5 M. The atom type is indicated on each panel. Colours blue$\rightarrow$red indicate increasing graphite surface charge densities from $|\sigma | =0 \rightarrow 0.77\; e$~nm$^{-2}$, with data on the left and right of each panel pertaining to simulations with positive and negative applied surface charges, respectively.}
    \label{fig:d5M}
\end{figure}

\begin{figure}[H]
    \centering
    \includegraphics[width=1\linewidth]{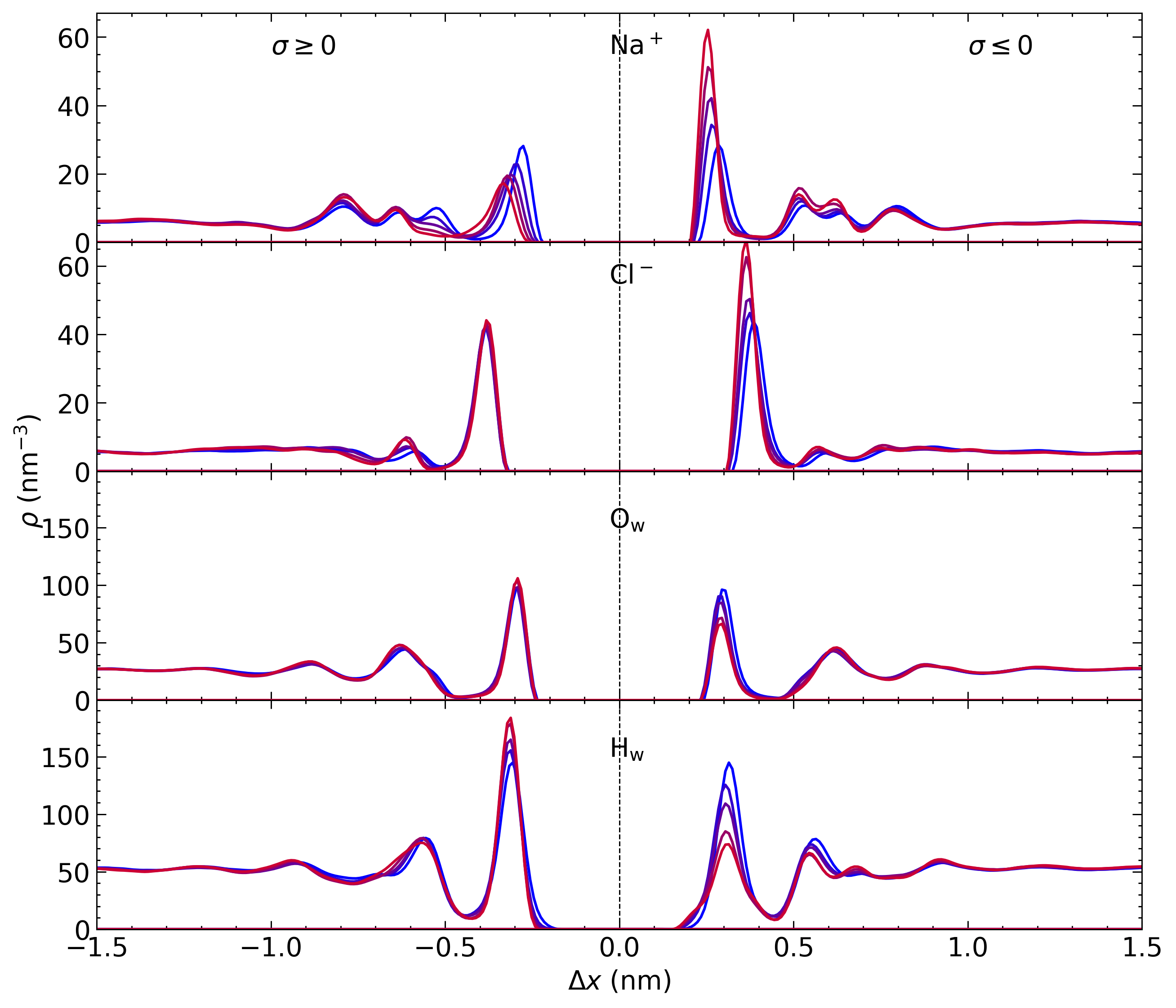}
    \caption{Water atom and ion solution densities, $\rho$, in the steady state determined as a function of distance from the graphite electrode, $\Delta x$, where the target bulk ion concentration was 10 M. The atom type is indicated on each panel. Colours blue$\rightarrow$red indicate increasing graphite surface charge densities from $|\sigma | =0 \rightarrow 0.77\; e$~nm$^{-2}$, with data on the left and right of each panel pertaining to simulations with positive and negative applied surface charges, respectively.}
    \label{fig:d10M}
\end{figure}

\begin{figure}[H]
    \centering
    \includegraphics[width=1\linewidth]{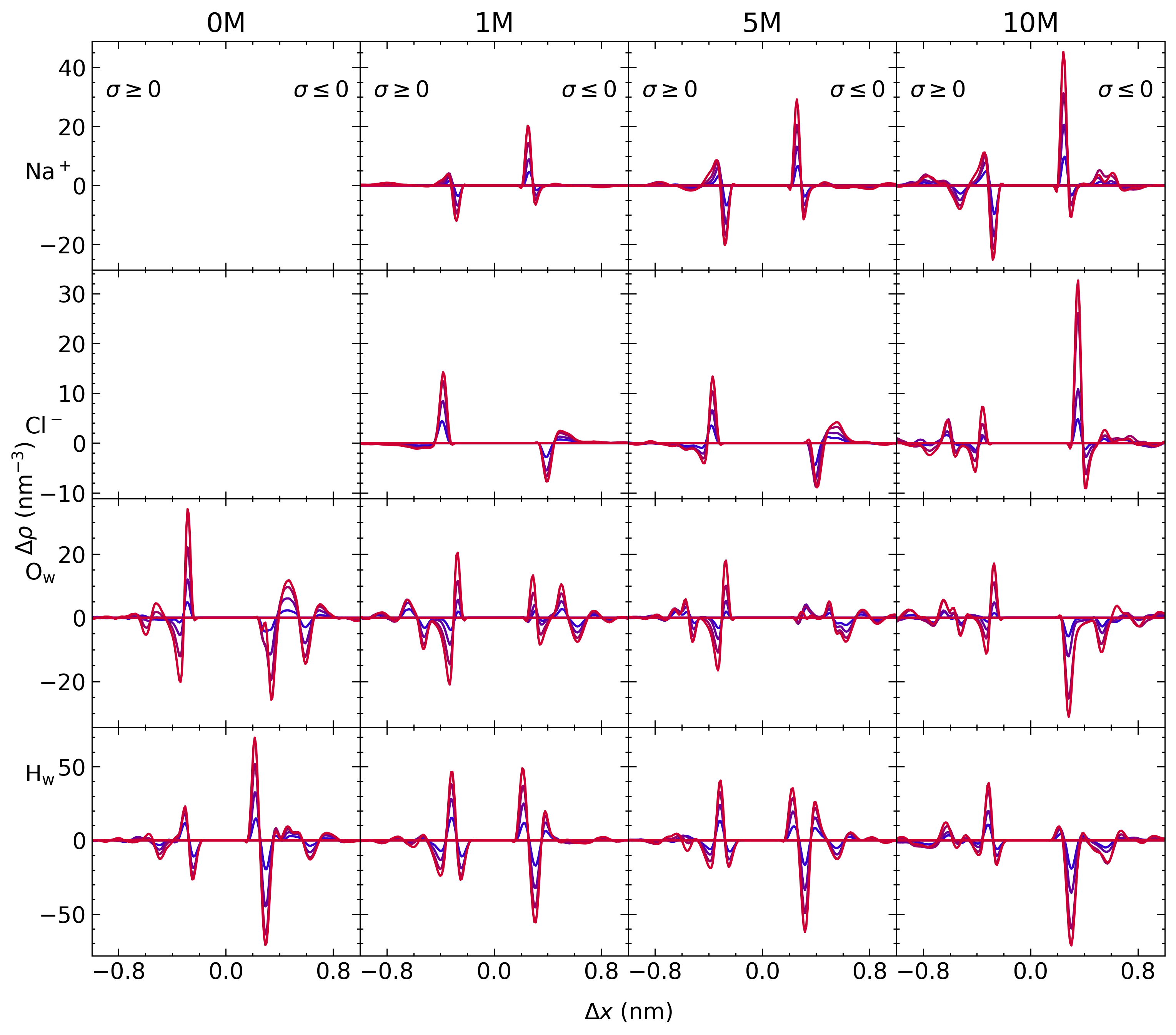}
    \caption{Change in the atom densities ($\Delta \rho$) as a function of distance from the graphite electrode according to the applied surface charge density ($|\sigma | =0 \rightarrow 0.77\; e$~nm$^{-2}$ as indicated by the blue$\rightarrow$red colour scale). Atom types are provided on the left, and target bulk concentrations are indicated the top of the grid.}
    \label{fig:densdiff}
\end{figure}

\begin{figure}[H]
    \centering
    \includegraphics[width=0.8\linewidth]{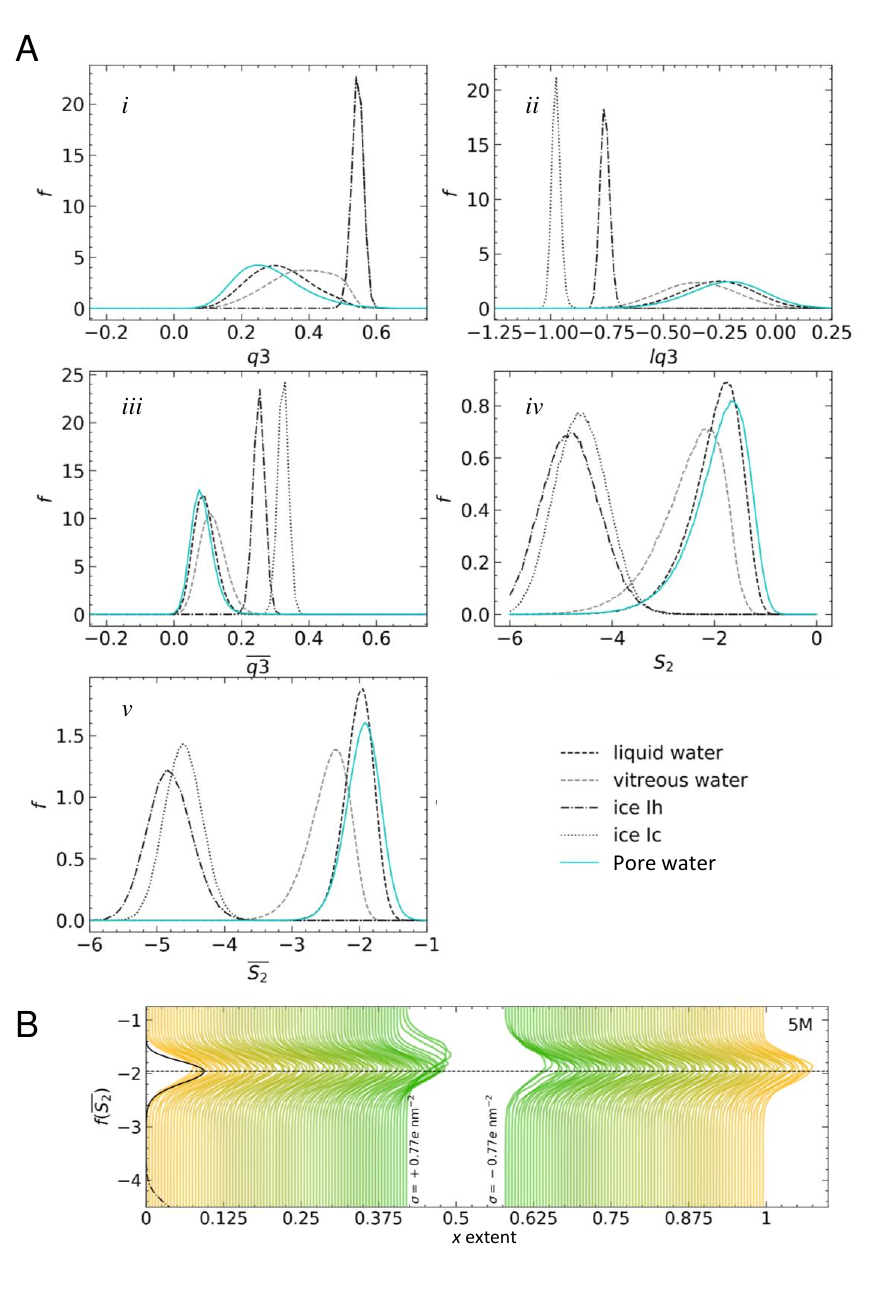}
    \caption{A) Order parameter analysis for water in C$\mu$MD simulations of 5 M NaCl(aq) at charged graphite with surface charge density, $|\sigma| = 0.77 \,e$ nm$^{-2}$. The probability density for $q3$ ($i$), local q3 ($ii; lq3$), local average q3 ($iii; \overline{q3}$), approximate pair entropy ($iv; S_2$) and local average approximate pair entropy ($v; \overline{S_2}$) are provided. The distributions for water in bulk liquid, vitreous (liquid water crash cooled to 100 K), hexagonal ice (ice $I_h$) and cubic ice ($I_c$) are also provided. B) Probability densities for $\overline{S_2}$ in slices of the cell $x$ axis. The bulk liquid water and ice $I_h$ distributions are provided on the $y$-axis.}
    \label{fig:SIOPs}
\end{figure}

\begin{figure}[H]
    \centering
    \includegraphics[width=1\linewidth]{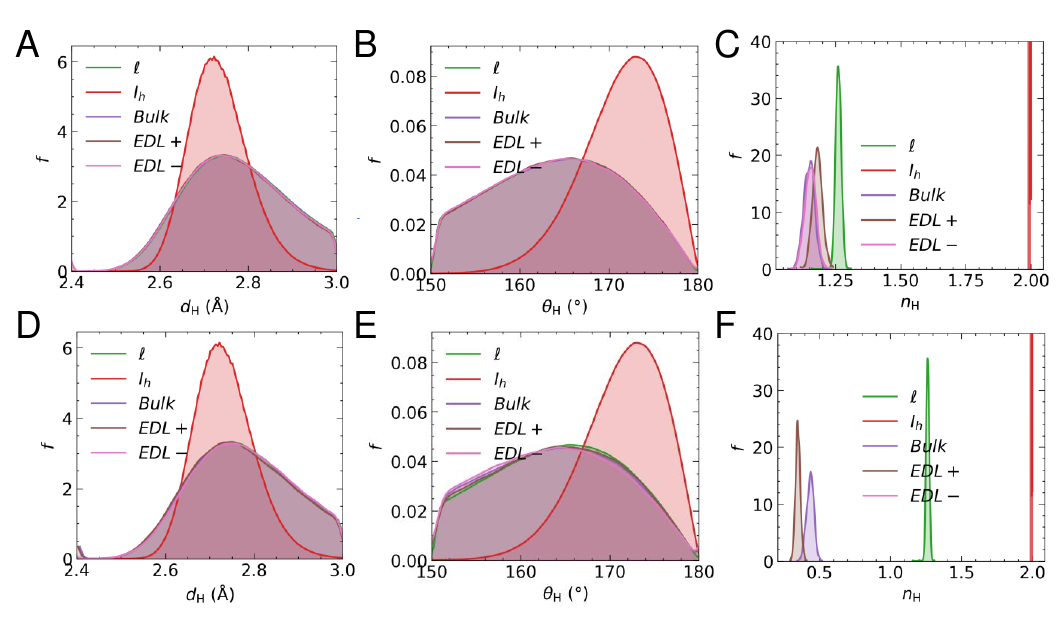}
    \caption{H-bonds for bulk liquid water ($\ell$), ice $I_h$, and water in the bulk and EDL regions of C$\mu$MD simulations of graphite in contact with NaCl(aq) where the surface charge density was $|\sigma|=0.77\,e$ nm$^{-2}$. $d_H$ and $\theta_H$ refer to the H-bond D---H$\cdots$A distance (A and D) and angle (B and E), respectively, where D, H and A refer to the oxygen donor, bonded hydrogen and oxygen acceptor, respectively. The number of H-bond donors per water molecule is provided in C, and F. A-C are for cases where the concentration was 0~M, while D-E provide the results from simulations at 10~M.}
    \label{fig:SIHbonds}
\end{figure}

\begin{figure}[H]
    \centering
    \includegraphics[width=0.45\linewidth]{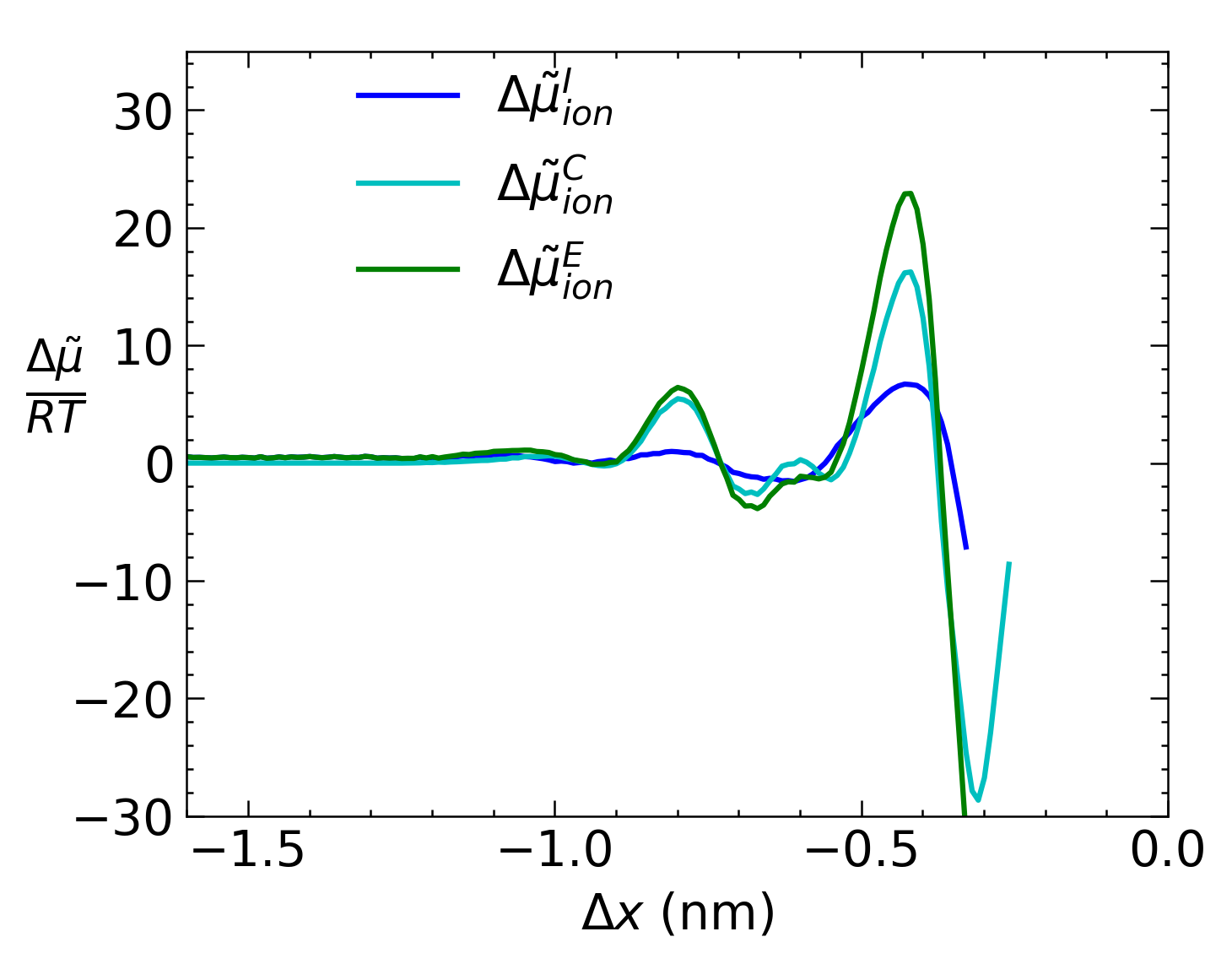}
    \caption{Contributions to the electrochemical potential of ions ($\Delta \tilde{\mu}$) in solution at graphite with $\sigma=+0.77 \; e$ nm$^{-2}$. $\Delta \tilde{\mu}^I=RT \left[ \ln m_{Na}(x) m_{Cl}(x) \right]$; $\Delta \tilde{\mu}^C=(2\omega -1)F \psi(x)$; and, $\Delta \tilde{\mu}^E=RT \ln \left[ \gamma_{ion}(m_{Na}(x)) \gamma_{ion}(m_{Cl}(x) ) \right]$.
    }
    \label{fig:cp-contributions}
\end{figure}

\begin{figure}[H]
    \centering
    \includegraphics[width=0.45\linewidth]{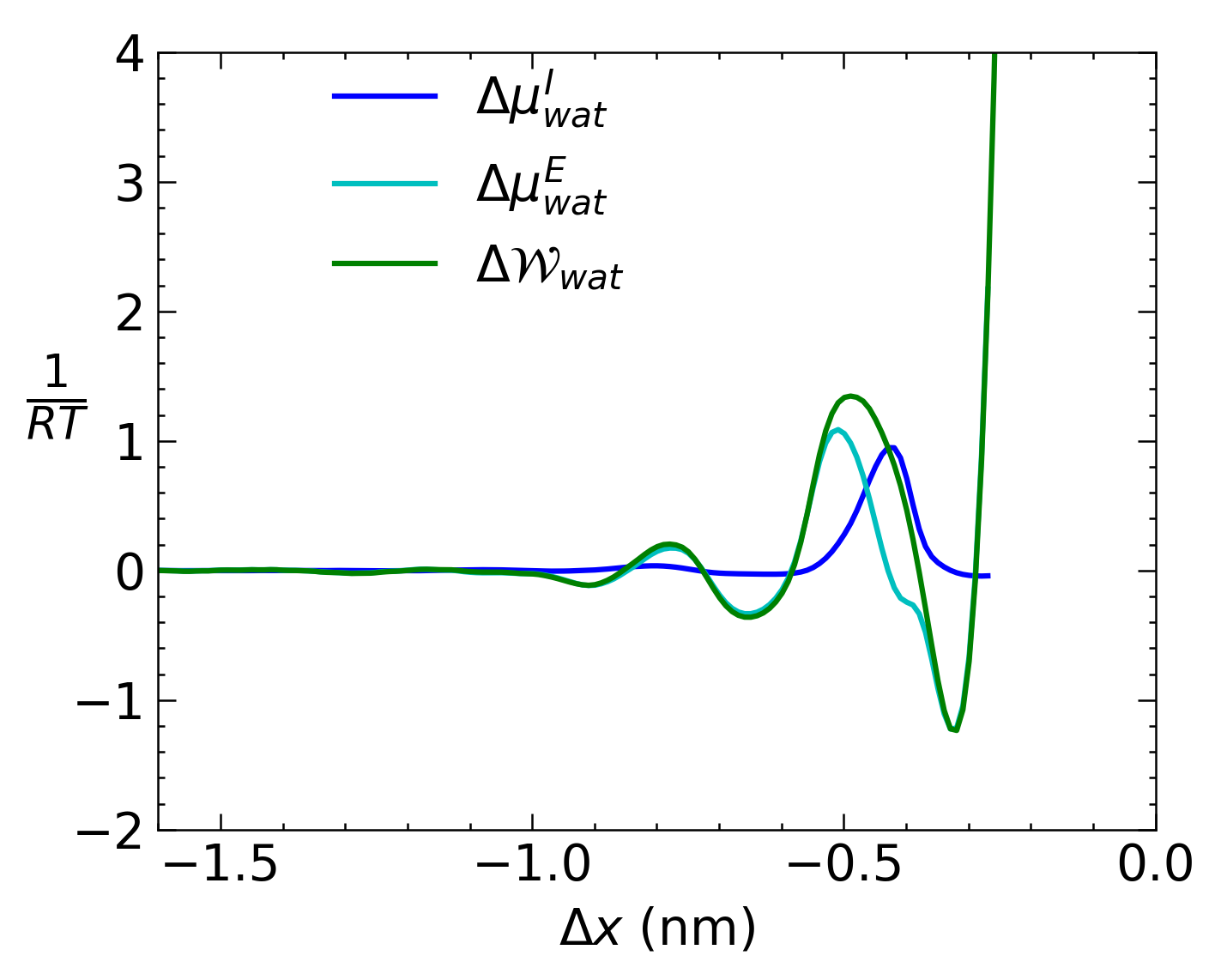}
    \caption{Contributions to ($\Delta \mathcal{W}_{wat}(x)$) in 1 M NaCl(aq) solution at graphite with $\sigma=+0.77e$ nm$^{-2}$. See equation \ref{eqn:fewat} in the main paper for details.
    }
    \label{fig:cp-wat-contributions}
\end{figure}

\begin{figure}[H]
    \centering
    \includegraphics[width=0.45\linewidth]{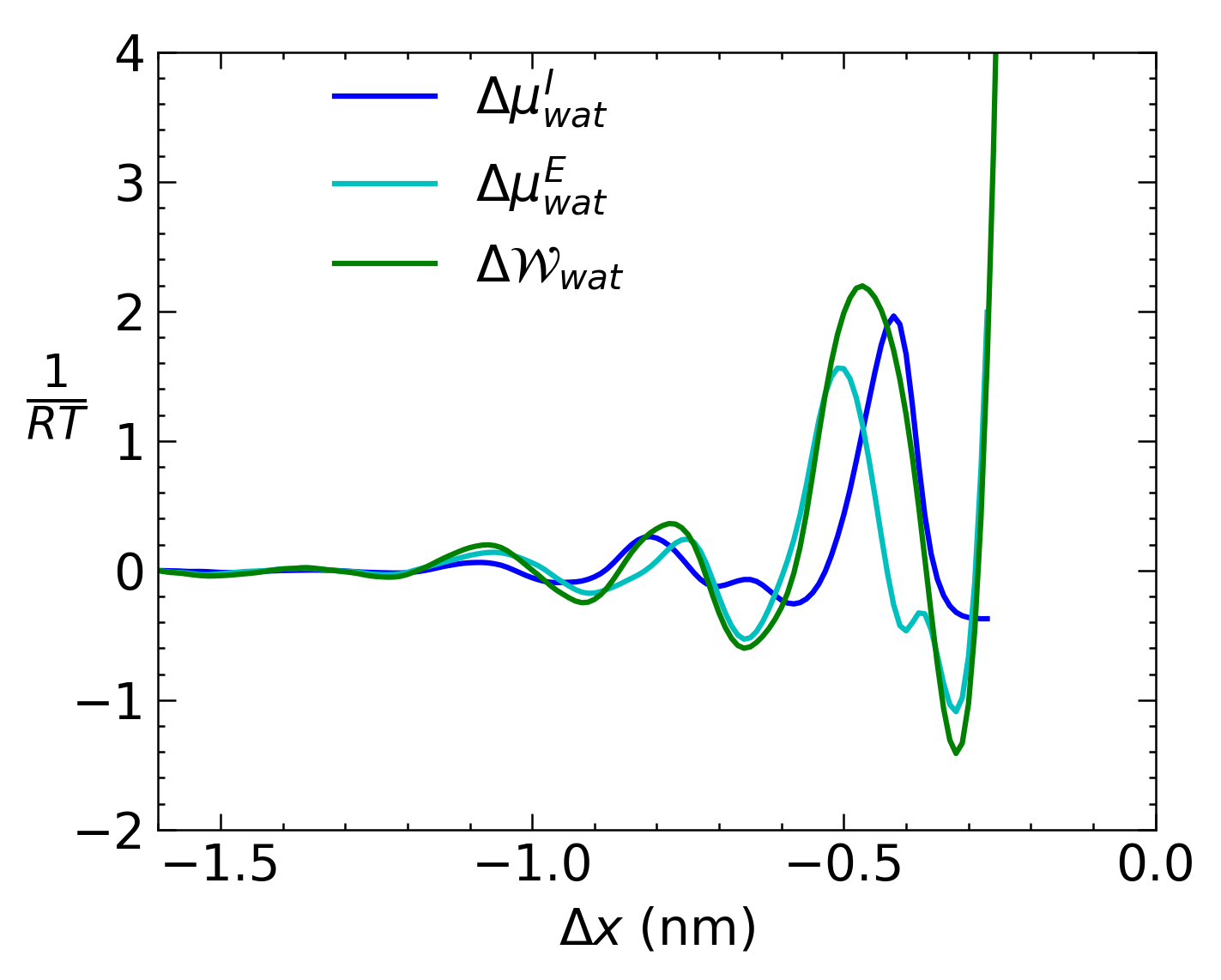}
    \caption{Contributions to ($\Delta \mathcal{W}_{wat}(x)$) in 10 M NaCl(aq) solution at graphite with $\sigma=+0.77e$ nm$^{-2}$. See equation \ref{eqn:fewat} in the main paper for details.
    }
    \label{fig:cp-wat-contributions-10M}
\end{figure}

\clearpage

\end{document}